\documentclass[letterpaper,11pt]{article}
\usepackage{jcappub} 

\usepackage{lineno}
\usepackage{fancyhdr}		

\usepackage{aas_macros}

\arxivnumber{2311.05586} 

\title{Pulsar Timing Anomalies:\\ {\huge A Window into Baryon Number Violation}}

\usepackage{siunitx}
\usepackage{amsmath}
\usepackage{mathtools}
\usepackage{graphicx}
\usepackage{wasysym}
\usepackage{amssymb}
\usepackage{array, multirow, bigdelim, makecell, booktabs} 
\usepackage{gensymb}
\usepackage{dcolumn}
\usepackage{bm}
\usepackage{slashed}
\usepackage{comment}
\usepackage[normalem]{ulem}

\usepackage{color}
\usepackage{float}
\usepackage{subfigure}
\usepackage[utf8]{inputenc}

\usepackage[dvipsnames]{xcolor}

\usepackage{erewhon}
\usepackage{lipsum}
\usepackage{lettrine}
\usepackage{GoudyIn}

\usepackage{subfigure}
\usepackage{float}

\newcommand{\ddotnu}{\ddot{\nu}}

\newcommand{\nobs}{n_{\text{obs}}}
\newcommand{\psrB}{PSR B0540--69}
\newcommand{\psrJ}{PSR J1640--4631}

\abstract{
We investigate the influence of a specific class of slow Baryon Number Violation (BNV)—one that induces quasi-equilibrium evolution—on pulsar spin characteristics. This work reveals how BNV can potentially alter observable parameters, including spin-down rates, the second derivative of spin frequency, and braking indices of pulsars. Moreover, we demonstrate that BNV could lead to anomalies in pulsar timing, along with a wide array of braking indices, both positive and negative. In addition, we examine the possibility of pulsar spin-up due to BNV, which may result in a novel mechanism for the revival of ``dead'' pulsars. We conclude by assessing the sensitivity required for future pulsar timing efforts to detect such BNV effects, thus highlighting the potential for pulsars to serve as laboratories for testing fundamental physics.
}

\author{Mohammadreza~Zakeri}
\affiliation{Department of Physics and Astronomy, University of Kentucky,\\ 
506 Library Drive, Lexington, KY~40506-0055, U.S.A.}

\emailAdd{M.Zakeri@uky.edu}

\usepackage{atbegshi,picture}
\AtBeginShipoutNext{\AtBeginShipoutUpperLeft{%
  \put(\dimexpr\paperwidth-1cm\relax,-1.5cm){\makebox[0pt][r]{\footnotesize \texttt{CETUP-2023-011}}}%
}}

\begin{document}
\maketitle

\flushbottom

\section{Introduction}
\label{sec:intro}

Pulsars serve as some of the universe's most precise clocks, offering a unique probe into the extraordinary conditions that prevail within them—conditions that cannot be replicated in any terrestrial laboratory~\cite{Weber:2006ep}. The precise timing of pulsar emissions has become a pivotal tool in the search for stochastic gravitational waves (GW)~\cite{Agazie_2023, EPTA:2023fyk, Reardon_2023, Xu_2023}. This study extends their utility into the domain of Baryon Number Violation (BNV)—a phenomenon integral to theoretical explanations for the baryon asymmetry of the universe~\cite{osti_4449128, Dine:2003ax, 1976ARA&A..14..339S, Zyla:2020zbs, Planck:2018vyg, 1981ApJ...246..557O}. We investigate how anomalies in pulsar timing may uncover subtle effects of BNV, broadening the scope of pulsars as laboratories for fundamental physics.

Despite the elusive nature of BNV in current terrestrial experiments, like those conducted at Super-Kamiokande~\cite{Super-Kamiokande:2015pys} and KamLAND~\cite{KamLAND:2005pen}, the extreme conditions within neutron stars provide a starkly different environment that may facilitate BNV processes. The intense gravitational forces within these astrophysical bodies can result in highly repulsive interactions among baryons, potentially enabling decay channels that are kinematically inaccessible in vacuum or within the confines of atomic nuclei~\cite{Berryman:2023rmh}. The dense cores of heavier neutron stars, possibly teeming with hyperons~\cite{1960SvA.....4..187A}, offer a natural laboratory for probing BNV within the strange quark sector. In this context, observational data from binary pulsars have been used~\cite{Berryman:2023rmh} to place constraints on hyperon decays to dark sectors~\cite{Alonso-Alvarez:2021oaj}, with implications that significantly extend beyond the reach of current terrestrial experiments~\cite{BESIII:2021slv}. These constraints are far more stringent, enhancing our understanding of BNV in these exotic environments.

The potential occurrence and rates of BNV across neutron stars, if they exist, are likely to be non-uniform and would be influenced by the complex interaction between particular particle physics models and the internal conditions of these stars. The composition of particles and their energy spectrum within a neutron star—factors critical to BNV rates~\cite{Berryman:2023rmh}—are primarily dictated by the star's total mass and the chosen equation of state (EoS). For instance, in scenarios anticipating BNV in the strange quark sector~\cite{Barducci:2018rlx, Heeck:2020nbq, Fajfer:2020tqf, Alonso-Alvarez:2021oaj}, BNV activity is contingent upon a critical mass threshold for hyperon presence in the core. Temperatures within neutron stars could also influence BNV, especially if BNV interactions are subject to the Pauli exclusion principle, which may block certain decay processes in cooler, denser states. This introduces a temperature dependence, enriching the BNV context but simultaneously complicating its detection; BNV effects could be notably diminished in the cold, degenerate matter prevalent in older neutron stars, making observational evidence contingent on accurate internal temperature estimations. Our approach is deliberately designed to be model-independent, enabling the examination of BNV effects across a diverse spectrum of neutron star masses and characteristics, providing a general framework for potential detection and analysis. 

In a recent study~\cite{universe10020067}, a broad range of scenarios were explored, detailing how dark sector dynamics and new physics can impact neutron stars. This present work narrows its focus, specifically examining how generic quasi-equilibrium BNV processes within neutron stars can be reflected in pulsar timing anomalies. An investigation of the interplay between BNV and neutron star dynamics introduced two pivotal assumptions to minimize model-specific dependencies~\cite{Berryman:2022zic}. The first assumption establishes that BNV proceeds at a slower rate than the chemical and hydrodynamic processes within a neutron star, ensuring quasi-equilibrium evolution. The second posits that there is no substantial accumulation of novel particles, other than those included in the standard Lagrangian defining the neutron star's structure, thereby preserving the configuration space delineated by the original EoS. Under these assumptions, BNV is envisioned to steer a neutron star through a series of quasi-equilibrium states, each described by the fixed baryon-conserving EoS. Such a transformation impacts the binary orbital periods of pulsars and their spin-down rates~\cite{Goldman:2009th, Goldman:2019dbq, Berryman:2022zic, Berryman:2023rmh}. Our previous paper laid the groundwork by exploring how neutron stars can generically constrain ongoing BNV, leveraging effects on binary pulsars' orbital periods~\cite{Berryman:2022zic}. Subsequent research delved deeper into the specifics of particle physics models and the role of dense matter in BNV processes within neutron stars~\cite{Berryman:2023rmh}. While these studies primarily focused on mechanisms affecting the orbital period of binary pulsars, the intricate details of BNV's influence on pulsar spin-down characteristics remained unexplored. This paper aims to bridge this gap by integrating pulsar rotation as a dynamic variable within the quasi-equilibrium BNV framework, thus offering a comprehensive analysis of its effects on various pulsar spin characteristics.

In section~\ref{sec:quasievol}, we lay the groundwork by developing the formalism for quasi-equilibrium evolution driven by BNV. Building on this foundation, section~\ref{sec:SD:formal} presents the derivation of the differential equation that governs the spin-down rate of pulsars under the influence of BNV. Subsequently, in sections~\ref{sec:pulsar:evol} and~\ref{sec:bridx}, we explore the implications of this framework on the evolution of pulsars, examining both the spin-down rates and the braking indices. The paper concludes with section~\ref{sec:obsprosp}, where we discuss the observational prospects for detecting BNV effects through precision measurements of pulsar spin frequency derivatives.

\section{Quasi-Equilibrium BNV} \label{sec:quasievol}
Equilibrium configurations of rotating cold neutron stars for a given choice of EoS are uniquely determined by two parameters: (1) central energy density (${\cal E}_c \equiv {\cal E}(r=0)$), and (2) central value for angular velocity of the local inertial frames ($\omega_c\equiv \omega(r=0)$)~\cite{Hartle:1967he, Glendenning:1997wn}. 
Here, $\omega(r)$ represents the rotational angular velocity of a freely falling particle around the center of the star as measured by an observer at infinity who is at rest relative to the star. The quasi-equilibrium evolution of an observable ($\mathcal{O}$) along this two-dimensional parameter space is given by
\begin{equation}
    \dot{\mathcal{O}} \equiv \frac{d \mathcal{O}}{dt} = \left(\frac{d {\cal E}_c}{dt}\right) \frac{\partial \mathcal{O}}{\partial {\cal E}_c} + \left(\frac{d \omega_c}{dt}\right) \frac{\partial \mathcal{O}}{\partial \omega_c}, \label{eq:Odot:general}
\end{equation}
where the baryon-number-conserving ($\dot{B} = 0$) trajectory is defined by
\begin{equation}
    \dot{{\cal E}}_c\, \partial_{{\cal E}} B = - \dot{\omega}_c\, \partial_{\omega} B,
\end{equation}
with the definitions $\dot{\omega}_c \equiv d{\omega}_c/dt$, $\dot{{\cal E}}_c \equiv d{{\cal E}}_c/dt$, $\partial_{{\cal E}} \equiv \partial / \partial_{{\cal E}_c}$, and $\partial_{\omega} \equiv \partial / \partial_{\omega_c}$. In the general case where $\dot{B} \neq 0$, and considering observed spin-down rates ($\dot{\Omega}$), we obtain
\begin{align}
    \dot{{\cal E}}_c =& \frac{\dot{B}\, \partial_{\omega} \Omega - \dot{\Omega} \, \partial_{\omega} B}{\partial_{{\cal E}} B\, \partial_{\omega} \Omega - \partial_{{\cal E}} \Omega \, \partial_{\omega} B},\\
 \dot{\omega}_c =& \frac{\dot{\Omega}\, \partial_{{\cal E}} B - \dot{B} \, \partial_{{\cal E}} \Omega}{\partial_{{\cal E}} B\, \partial_{\omega} \Omega - \partial_{{\cal E}} \Omega \, \partial_{\omega} B}.
\end{align}
This leads to the following form for Eq.~\eqref{eq:Odot:general}:
\begin{equation}
    \left[\dot{\mathcal{O}} \right]_{\rm BNV} = \frac{ \left[ \partial_{\omega} \Omega\, \partial_{{\cal E}} \mathcal{O} - \partial_{{\cal E}} \Omega\, \partial_{\omega} \mathcal{O} \right] \dot{B}
    +
    \left[ \partial_{{\cal E}} B\, \partial_{\omega} \mathcal{O}  - \partial_{\omega} B\, \partial_{{\cal E}} \mathcal{O} \right] \dot{\Omega} 
    }{\partial_{{\cal E}} B\, \partial_{\omega} \Omega - \partial_{{\cal E}} \Omega \, \partial_{\omega} B} \equiv \beta({\mathcal{O}})\, \dot{B} + \gamma({\mathcal{O}})\, \dot{\Omega},\label{eq:Odot:general:2}
\end{equation}
where the operators $\beta$ and $\gamma$ correspond to $\left(\partial / \partial B \right)_{\Omega}$ and  $\left(\partial / \partial \Omega \right)_{B}$, respectively. We introduce dimensionless parameters $b$ and $g$ via
\begin{align}
    b (\mathcal{O}) \equiv& \beta(\mathcal{O}) \times \frac{B}{\mathcal{O}}, \\
    g (\mathcal{O}) \equiv& \gamma(\mathcal{O}) \times \frac{\Omega}{\mathcal{O}},
    \label{eq:defs:b:g}
\end{align}
to represent the relative rate of change in an observable divided by the relative rate of change in the total baryon number and angular velocity, respectively.

We now demonstrate the derivation of constraints on the order of magnitude of the BNV rate in individual pulsars, which can be determined independently of the detailed mechanisms of spin-down formalism, as detailed in Eq.~\eqref{eq:bnv_lim}. The observed spin-down rate ($\dot{P}_{\rm obs}$) of a pulsar is a combination of both intrinsic ($\dot{P}_{\rm int}$) and extrinsic ($\dot{P}_{\rm ext}$) terms. Extrinsic factors encompass elements such as radial acceleration and the Shklovskii effect, which is a consequence of the pulsar's proper motion and its radial velocity with respect to the observer~\cite{1970SvA....13..562S}. The BNV effects would contribute to $\dot{P}_{\rm int}$. To discern this intrinsic spin-down rate, one needs to subtract the extrinsic contributions from the observed rate: 
 \begin{equation}
\begin{split}
   \left(\frac{\dot{P}}{P}\right)_{\rm obs} - \left(\frac{\dot{P}}{P}\right)_{\rm ext} = \left(\frac{\dot{P}}{P}\right)_{\rm int} =& \overbrace{-\frac{\dot{J}_{\rm EM}}{J} + \left(\frac{\partial I}{\partial \Omega}\right)_B \left(\frac{\dot{\Omega}}{I}\right)}^{\textrm{EM}}
   - \frac{\dot{J}_{\rm BNV}}{J} + \left(\frac{\partial I}{\partial B}\right)_{\Omega} \left(\frac{\dot{B}}{I}\right), \\
    =& \left(\frac{\dot{P}}{P}\right)_{\rm EM} - \frac{\dot{J}_{\rm BNV}}{J} - b(I) \times \Gamma.
\end{split}
\end{equation}
In this equation, ``EM'' represents the standard electromagnetic contributions, and $J$ denotes the pulsar's total angular momentum relative to its center of mass. The BNV rate is symbolized by $\Gamma \equiv -\dot{B}/B$. One critical factor in constraining $\Gamma$ from observations is a comprehensive understanding of both the electromagnetic model and the magnetic field. A challenge here is that magnetic fields are usually inferred from the observed spin-down rate. Direct measurements are typically made possible in specific mass-accreting neutron stars, where X-ray observations can detect electron cyclotron resonances~\cite{10.1143/PTPS.151.54}.

As we will show shortly, we are interested in scenarios where the term $(\dot{J}_{\rm BNV}/J)$ is negligible. With the assumption that $(\dot{P}/P)_{\rm EM} > 0$, it is evident that the EM and BNV terms adopt identical (or opposite) signs based on whether $b(I)$ is negative (or positive). Considering the cases where $b(I) < 0$, specifically for pulsars approaching the maximum mass ($M_{\rm TOV} - M \lessapprox 0.1 M_{\odot}$ as illustrated in figure~\ref{fig:b-factor}), we can derive a conservative limit on $\Gamma$. This limit requires that the BNV contribution remains below the entirety of the intrinsic spin-down rate, leading to $-b(I)\times \Gamma < (\dot{P}/P)_{\rm int}$. On the contrary, for lighter pulsars where $b(I) > 0$, one could argue two potential scenarios:

\begin{enumerate}
    \item $b(I) \times \Gamma \lessapprox (\dot{P}/P)_{\rm int}$, or
    \item A cancellation occurs between EM and BNV contributions. This accidental cancellation is perplexing, given the diverse origins and independent evolutionary paths of the EM and BNV terms. In fact, the likelihood of such an occurrence manifesting across an entire pulsar population is significantly low.
\end{enumerate}

Given these considerations, we can set constraints on BNV through pulsar spin-down observations:

\begin{equation}
\left|b(I) \times \Gamma \right| \lessapprox \left(\frac{\dot{P}}{P}\right)_{\rm int}, \qquad \text{assuming} \quad \frac{\dot{J}_{\rm BNV}}{J} \ll b(I) \times \Gamma. \label{eq:bnv_lim}
\end{equation}

We now demonstrate how this limit can be applied across an entire population of pulsars. Assuming BNV is density-driven\footnote{This implies that the BNV rate increases monotonically with the density of the neutron star medium, for instance, through hyperon decays.} and is not influenced by other variables, like core temperature, it would be expected to occur more rapidly in more massive pulsars.
In this work, we refer to this type of BNV as ``universal.''
Applying the limit in Eq.~\eqref{eq:bnv_lim} to the notably heavy pulsar, PSR J0348$+$0432~\cite{Antoniadis:2013pzd} with a mass $M \approx 2 M_{\odot}$, provides an upper bound on universal BNV: $\Gamma \lesssim 2 \times 10^{-10}\, {\rm yr}^{-1}$. Exceeding this rate would likely manifest through distinct spin-down characteristics observable in the most massive pulsars, particularly as these effects become more pronounced in older millisecond pulsars with smaller spin-down rates ($\dot{P}$). Therefore, we infer that a universal BNV rate exceeding $\sim 10^{-10}\, {\rm yr}^{-1}$ across the pulsar population, especially in those lighter than $2 M_{\odot}$, is highly improbable. The subsequent sections will explore the effects of both universal and non-universal BNV on pulsar spin-down rates and braking indices in greater detail.

\section{Spin-Down Formalism}
\label{sec:SD:formal}

In this section, we aim to derive an equation that describes the rotational dynamics of pulsars, accounting for both BNV and EM radiation effects. The rate of change in the pulsar's rotational energy ($\dot{E}$) can be expressed as $\dot{E} = \dot{E}_{\rm BNV} + \dot{E}_{\rm EM}$, in which $\dot{E}_{\rm BNV}$ is the rotational energy loss due to BNV, in the absence of EM radiations, and $\dot{E}_{\rm EM}$ is due to EM radiation, in the absence of BNV. We first establish a relationship between the spin down rate ($\dot{\Omega}_{\rm BNV}$) and the changes in the moment of inertia ($\dot{I}_{\rm BNV}$) by neglecting the EM contribution. We describe this in terms of angular momentum loss as follows: $\dot{J}_{\rm BNV} = \dot{I}_{\rm BNV} \Omega + I \dot{\Omega}_{\rm BNV}$. Let us define a dimensionless quantity $\epsilon\equiv (\dot{J}_{\rm BNV}/J)/(\dot{I}_{\rm BNV}/I)$ to parameterize the angular momentum loss such that
\begin{equation}
\left[\frac{\dot{\Omega}}{\Omega}\right]_{\rm BNV} = \frac{b(I)\, \left(\epsilon - 1\right)}{1 - \left(\epsilon - 1\right) g(I)} \left(\frac{\dot{B}}{B}\right), \label{eq:I:omega:general}
\end{equation}
in which we identify the case of purely radial matter redistribution with $\epsilon=0$, and the case of purely isotropic (in the fluid's rest frame) energy loss with $\epsilon=1$. The rate of change in the rotational energy induced by BNV is
\begin{equation}
    \left[-\dot{E}\right]_{\rm BNV} = \left[-\frac{d(I\Omega^2/2)}{dt}\right]_{\rm BNV} = \frac{I\Omega^2}{2} \left(\frac{(1-2\epsilon)\, b(I)}{1 + (1-\epsilon)g(I)}\right) \left(\frac{\dot{B}}{B}\right).
\end{equation}
Having established the role of BNV, let us now turn our attention to the impact of magnetic braking. In the original magnetic dipole radiation model~\cite{1968Natur.219..145P, 1969Natur.221..454G}, the energy loss term ($\dot{E}$) is set equal to the emission power of a magnetic dipole in vacuum. However, it was shown later~\cite{1969ApJ...157..869G} that the vacuum solution around a pulsar is unstable in the sense that particles are pulled off from the pulsar surface into its magnetosphere and the plasma effects may not be ignored. We use a dipole (D) model fitted by a magnetosphere simulation~\cite{Philippov:2013aha} which characterizes the energy loss as
\begin{equation}
    \left[-\dot{E}\right]_{\rm D} = \left[-\frac{d(I\Omega^2/2)}{dt}\right]_{\rm D} = R^6 H^2 \Omega^4 \left[\kappa_0 + \kappa_1 \sin^2(\alpha)\right], \label{eq:Edot:dipole}
\end{equation}
in which $H$ is the poloidal dipole magnetic field, $\kappa_0 \approx 1$, $\kappa_1 \approx 1.2$, and $\alpha$ is the angle between the rotation and the dipole axes which evolves as
\begin{equation}
    \dot{\alpha} = -\kappa_2 \left(\frac{R^6 H^2\Omega^2}{I}\right) \sin(\alpha) \cos(\alpha), \label{eq:alpha}
\end{equation}
in which $\kappa_2 \approx 1$. The evolution timescale for $\alpha$ in oblique isolated pulsars is estimated from polarization measurements of radio pulsars to be about $10^7$ year~\cite{10.1046/j.1365-8711.1998.01369.x}. 

We now combine the dipole (D) and BNV terms to arrive at
\begin{equation}
\begin{split}
    -\dot{E} = -\left[\dot{E}\right]_{\rm BNV} - \left[\dot{E}\right]_{\rm D} &= \frac{I\Omega^2}{2} \left(\frac{(1-2\epsilon)\, b(I)}{1 + (1-\epsilon)g(I)}\right) \left(\frac{\dot{B}}{B}\right) +  R^6 H^2 \Omega^4 \left[\kappa_0 + \kappa_1 \sin^2(\alpha)\right] \\
    &= -\frac{I \Omega^2}{2} \left[b(I) \left(\frac{\dot{B}}{B}\right) + g(I) \left(\frac{\dot{\Omega}}{\Omega}\right) \right] - I \Omega \dot{\Omega},\label{eq:Eloss}
\end{split}
\end{equation}
in which the last equality follows from $E = I \Omega^2 / 2$ and Eq.~\eqref{eq:Odot:general:2} with $\mathcal{O} = I$. We rearrange the terms in Eq.~\eqref{eq:Eloss} to turn it into a differential equation governing the spin-down rate:
\begin{equation}
b(I)\left( \frac{(1-\epsilon)[2+g(I)]}{1 + (1-\epsilon) g(I)}\right) I \Omega \left(\frac{\dot{B}}{B}\right)  + \left[ 2 + g(I) \right] I \dot{\Omega} +  2R^6 H^2 \Omega^3 \left[\kappa_0 + \kappa_1 \sin^2(\alpha)\right]  = 0. \label{eq:SD}
\end{equation}
Equation~\eqref{eq:SD} thus encapsulates the combined effects of BNV and EM radiation on the spin-down rate of pulsars. We note that the gravitational radiation contribution $\dot{E}_{\rm GW} \propto \Omega^6$~\cite{shapiro2008black} is subdominant.

The role of BNV in the angular momentum loss of pulsars is multifaceted, especially concerning the emission of specific particles. The BNV reaction(s) can produce particles that have negligible interactions with the SM sector. We refer to these as ``dark particles'' and denote them by $\chi$. If $\chi$ has a sufficiently low mass, it could potentially escape the gravitational field of a pulsar, subsequently reducing its angular momentum. In our quasi-equilibrium BNV formalism, $\chi$ can:
\begin{enumerate}
\item Escape the pulsar.
\item Convert into other dark particles that ultimately leave the star.
\item Revert into SM particles.
\end{enumerate}
Additionally, neutrinos are emitted via Urca type processes. These reactions may be enhanced due to the beta disequilibrium induced by BNV~\cite{Berryman:2022zic}. A combination of these scenarios is also plausible. We note that if the dark sector is isolated from the SM, the angular momentum loss in the first two scenarios would be equal. In the third scenario, this loss would result from neutrino emission by Urca reactions. 

Working within the Newtonian approximation, we estimate the angular momentum loss ($\epsilon$) due to emission of $\chi$ and neutrinos separately, with the assumption that these emissions are isotropic in the fluid rest frame. For simplicity, we have adopted the natural units system, where the speed of light ($c$) is set to $1$.

\begin{figure}[t]
    \centering
    \includegraphics[width=0.95\textwidth]{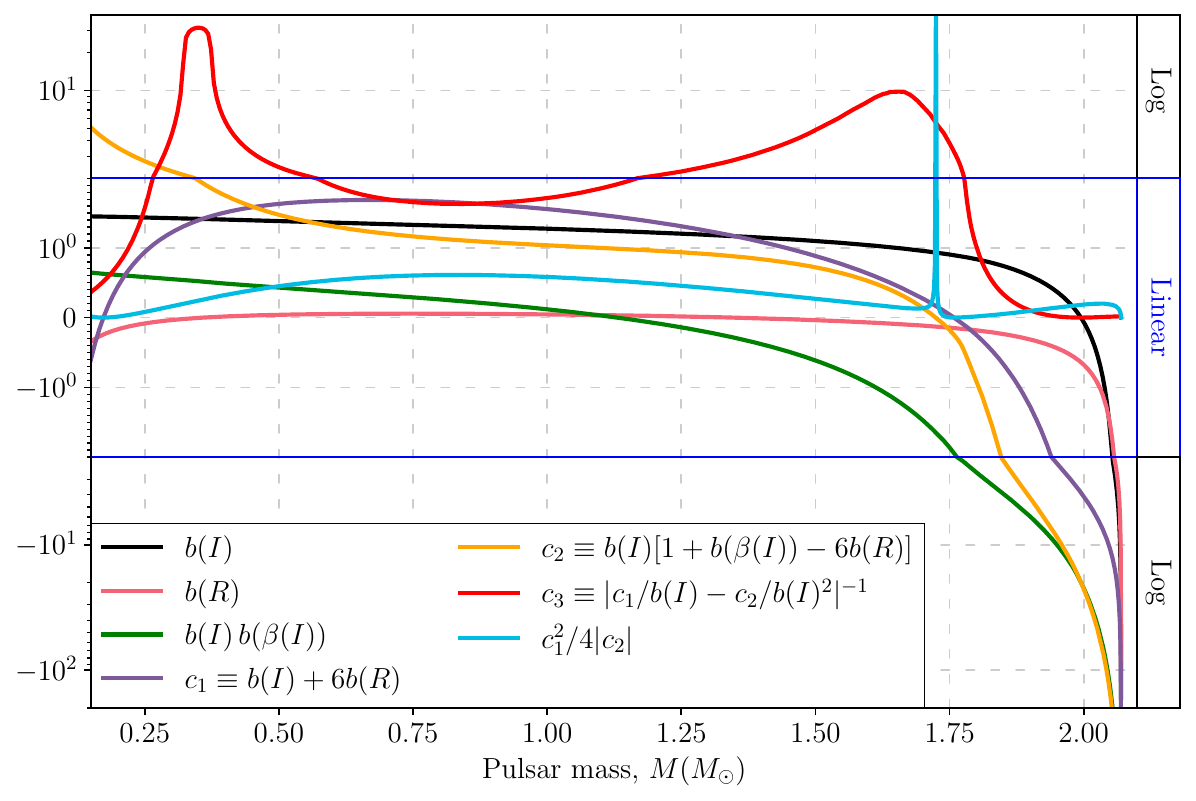}
    \caption{
    Variation of the pertinent coefficients for calculating the braking index ($n$) in Eq.~\eqref{eq:n:slow}, and the second derivative of the spin frequency ($\ddot{\nu}$) in Eq.~\eqref{eq:nuddot}, as functions of pulsar mass, assuming the DS(CMF)-1 EoS~\cite{Dexheimer:2008ax}. Coefficient $c_3$ is utilized to distinguish BNV influences from extrinsic contributions, as described in Eq.~\eqref{eq:bnv_vs_Shk} to Eq.~\eqref{eq:bnv_vs_jerk}. The y-axis employs symmetrical logarithmic scaling: it is linear within the interval \([-2, 2]\)—indicated by the solid blue horizontal lines—and logarithmic beyond this range.}
    \label{fig:b-factor}
\end{figure}

A dark particle ($\chi$), produced with an energy $E_{\chi}$ and at a distance $r_{\perp} = r \sin(\theta)$ from the rotation axis, carries an angular momentum $\delta J_{\chi} \approx E_{\chi} r_{\perp}^2 \Omega$~\cite{1977Natur.267..501K, Baumgarte_1998}. For a neutron star rotating uniformly, the dimensionless parameter $\epsilon_{\chi}$, can be approximated by
\begin{equation}
\epsilon_{\chi} \approx \frac{-\left[\int d^3x\, r_{\perp}^2 \, Q_{\chi}(r)\right] / I}{b(I) \left(\dot{B}/B\right) + g(I)\left(\dot{\Omega}/\Omega\right)} \approx \frac{\overline{E}_{\chi} / (M/B)}{b(I) + g(I)\left(\dot{\Omega}/\Omega\right)\left(B/\dot{B}\right) } \left(\frac{R_{\chi}}{R}\right)^5, \label{eq:ang_loss:chi:general}
\end{equation}
in which we assumed that the decays are active within $R < R_{\chi}$, $M$ is the neutron star mass, $Q_{\chi}(r) = \Gamma(r)\times n(r) \times \overline{E}_{\chi}(r)$ is the emissivity rate of $\chi$, $n(r)$ is the local baryon number density, and $\Gamma(r)$ is the BNV rate per baryon, which if we take to be radially uniform is given by $\Gamma (r) = \Gamma = -\dot{B}/B$. Note that when only a specific type of baryon ($i$) is assumed to decay, we have $\Gamma = f_i\, \Gamma_i$, in which $f_i \equiv B_i/B$ is the fraction of species $i$, and $\Gamma_i \equiv \dot{B}_i/B_i$ is the decay rate of baryon species $i$. Given that $g(I)>0$, the two terms in the denominator of Eq.~\eqref{eq:ang_loss:chi:general} have the same sign (with $b(I)\sim 1$ from figure~\ref{fig:b-factor}) for spinning-down ($\dot{\Omega}<0$) pulsars, except for neutron stars with masses very close to the maximum mass ($M_{\rm TOV}$). Moreover, the inclusion of spin-down contributions is particularly relevant for pulsars with exceptionally high rotational speeds~\cite{Hamil:2015hqa}. Therefore, a conservative limit on $\epsilon_{\chi}$ in Eq.~\eqref{eq:ang_loss:chi:general} is given by $\approx\left(\overline{E}_{\chi} / {\rm GeV}\right)(R_{\chi}/R)^5$, which would be less than $1\%$ for $R_{\chi} < 0.4 R$. We focus on scenarios where the BNV decays would be mainly occurring at the core of neutron stars such that $\epsilon_{\chi} \ll 1$.

We now estimate the angular momentum loss due to indirectly produced (Urca) neutrinos ($\epsilon_{\nu}$). This process is particularly significant because it serves as a primary cooling mechanism for the star through neutrino emission. The direct Urca process, described as $n \to p + \ell + \bar{\nu}_{\ell}$ and its inverse~\cite{gamow1970my}, is only feasible in the core of neutron stars above a certain threshold mass and is forbidden in lighter stars due to energy and momentum conservation constraints~\cite{BOGUTA1981255, PhysRevLett.66.2701}. For stars not meeting the conditions for direct Urca, a modified channel exists: the modified Urca process involves a spectator nucleon ($N$) and is represented by $N + n \to N + p + \ell + \bar{\nu}_{\ell}$ and its inverse~\cite{PhysRevLett.12.413, PhysRev.140.B1452}. The neutrino emissivity ($Q_{\nu}$) from the modified Urca process is approximately $\approx 3 \times 10^{21}\, T_9^8\, {\rm erg}\, {\rm s}^{-1}\, {\rm cm}^{-3}$, with $T_9 \equiv T / 10^9\, {\rm K}$ representing the star's internal temperature normalized to a billion Kelvin~\cite{PhysRevLett.66.2701}. Consequently, the relative angular momentum loss rate due to neutrinos, denoted by $\epsilon_{\nu}$, is estimated to be:
\begin{equation}
\epsilon_{\nu} \approx \frac{2.7 \times 10^{-7}\, {\rm yr}^{-1}}{b(I) \left(\dot{B}/B\right) + g(I)\left(\dot{\Omega}/\Omega\right)}  T_9^8,
\end{equation}
For neutron stars with BNV rates on the order of $\dot{B}/B \sim 10^{-10}\, {\rm yr}$ and internal temperatures less than $2 \times 10^{8}\, K$, $\epsilon_{\nu}$ is less than $1\%$. We are interested in BNV effects within the cooler and older neutron stars, where $T_9 \ll 1$, further ensuring that $\epsilon_{\nu}$ remains much less than one.

Given that the angular momentum loss attributed to dark particles and neutrinos is negligible for the cases we consider, we proceed by setting $\epsilon=0$ for the remainder of this paper. We rewrite Eq.~\eqref{eq:SD} in a more convenient manner using Eq.~\eqref{eq:defs:b:g} and the following definitions

\begin{align}
\mu_D \equiv& R^3 H, \quad\quad &\text{(magnetic dipole moment)}, \\
\delta \equiv& \frac{\Gamma\, I}{\mu_D^2\, \Omega^2\, \left[\kappa_0 + \kappa_1 \sin^2(\alpha)\right]}, \quad \quad &\text{(BNV to dipole spin-down rate ratio)}.
\label{eq:defs}
\end{align}
The parameter $\delta$ quantifies the relative influence of BNV compared to the magnetic dipole mechanism in the pulsar's spin-down process. With these new definitions, Eq.~\eqref{eq:SD} is recast into a more tractable form:
\begin{equation}
\dot{\Omega} = \frac{\mu_D^2 \Omega^3}{I} \left[\kappa_0 + \kappa_1 \sin^2(\alpha)\right]\left( \frac{b(I)\, \delta}{1 + g(I)} -  \frac{2}{ 2 + g(I) } \right), \label{eq:SD:dimless}
\end{equation}
highlighting a structure reminiscent of phenomenological spin-down models that account for pulsar braking indices ranging between $1 < n < 3$. Specifically, adopting $\dot{\Omega} = - a_1 \Omega^3 - a_2 \Omega$, the linear term emerges in the particle flow model~\cite{1969ApJ...158..727M}, where $a_2$ equals $(\pi \Phi^2 / 4 I)$ and $\Phi$ represents the magnetic flux associated with the particle flow. On short timescales during which $b(I)$ and $g(I)$ can be taken to be approximately constant, this linear term would be analogous to the term $-b(I) \Gamma / [1+g(I)]$ in our framework .

Considering the effects of rotation on the static structure of neutron stars, we note that they contribute at an order $\sim \Omega^2 / \left(G\, M / R^3\right) \lesssim 0.4\, ({\rm ms}/P)^2$~\cite{Hartle:1967he}. For the known pulsar population, ranging from the slowest~\cite{Caleb:2022xyo} to the fastest~\cite{Hessels:2006ze}, this influence varies widely, between $10^{-10}$ and $0.2$. Our dimensional analysis suggests that $g(I)$ is less than $0.4$ for pulsar with periods slower than $1\, {\rm ms}$. Therefore, we do not expect a substantial effect from rotation, as described by $g(I)$ in Eq.~\eqref{eq:SD:dimless}. This accounts for why the observed deviations of some pulsar braking indices from the expected value of $n=3$ cannot be explained solely by the inclusion of $g(I)$, as indicated in Ref.~\cite{Hamil:2015hqa}. Similarly, the dependence of $b(I)$ on $\Omega$ is expected to be minor. The contribution from BNV would be larger than that from rotation effects if $b(I) \delta > g(I) / 2$. This condition translates to $\delta \gtrsim 0.05$ for $P \gtrsim 2$ (ms). In this case Eq.~\eqref{eq:SD:dimless} takes a much simpler form: 
\begin{equation}
\dot{\Omega} = \frac{\mu_D^2 \Omega^3}{I} \left[\kappa_0 + \kappa_1 \sin^2(\alpha)\right]\left( b(I)\, \delta - 1 \right). \label{eq:SD:dimless:slow}
\end{equation}

\begin{figure}[t]
    \centering
    \includegraphics[width=0.8\textwidth]{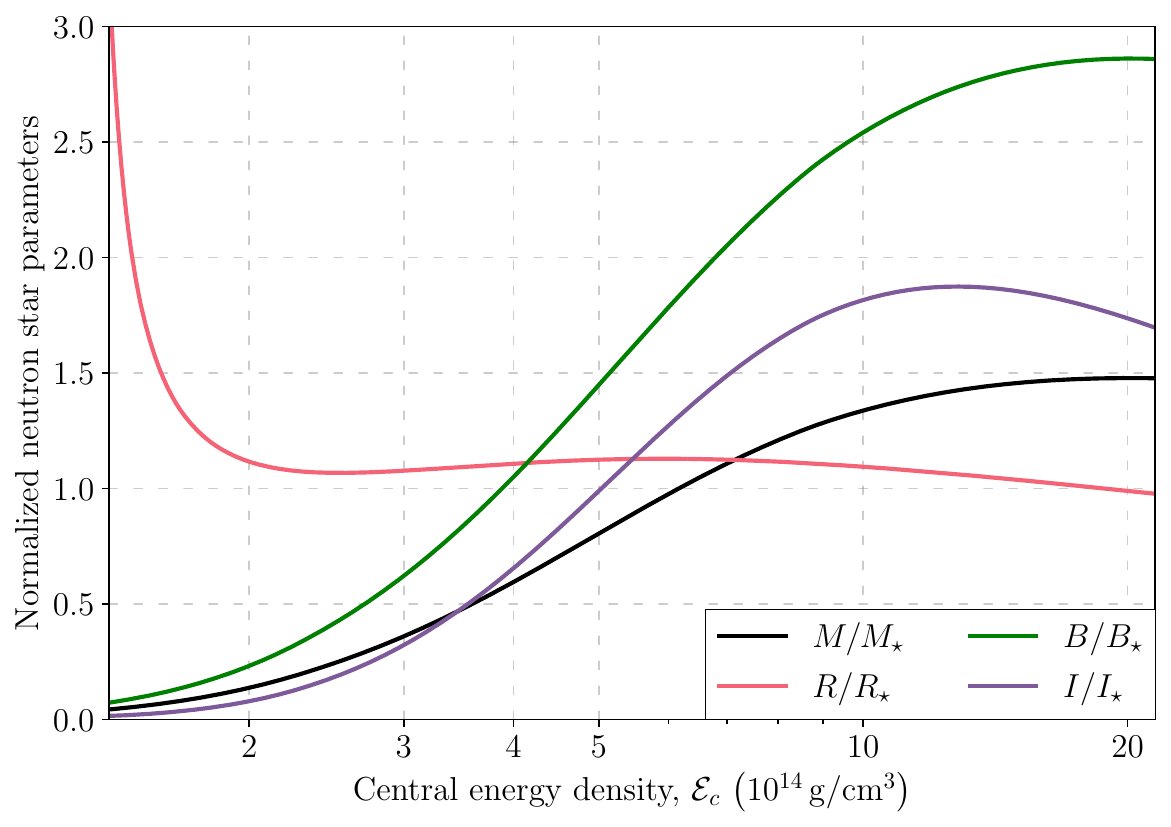}
    \caption{
    Variations of key neutron star properties as functions of the central energy density (${\cal E}_c$). Displayed quantities include the mass ($M$), radius ($R$), baryon number ($B$), and moment of inertia ($I$), each normalized to their typical values: $M_{\star} = 1.4\, M_{\odot}$, $R_{\star} = 12\, {\rm km}$, $B_{\star} = 10^{57}$, $I_{\star} = 70\, {\rm M_{\odot}\, {km}^2}$. This sequence is generated using the DS(CMF)-1 EoS~\cite{compose_CMF1}.}
    \label{fig:sequence}
\end{figure}

To calculate the b-factors crucial to our analysis, we generated a sequence of static, non-rotating neutron stars using the DS(CMF) EoS~\cite{Dexheimer:2008ax}—the first hadronic variant that includes a crust~\cite{Gulminelli:2015csa}, as documented in the CompOSE database~\cite{CompOSECoreTeam:2022ddl}. Detailed descriptions of the EoS, including its Lagrangians and particle constituents, are provided in Ref.~\cite{Berryman:2023rmh}. The resultant sequence is depicted in figure~\ref{fig:sequence}. We derived the b-factors associated with this sequence, and plotted them in figure~\ref{fig:b-factor}. This figure indicates that the dimensionless b-parameter associated with the moment of inertia, $b(I)$, remains close to $1$ for most pulsars, except near the maximum mass threshold $M_{\rm TOV}$. 

While the adoption of a different EoS could yield variations in specific mass values, the general behavior and trends of the b-factor curves are robust across different EoS choices. The pivotal mass values that we reference later in the text—those that signal shifts in the curves' behaviors—would indeed change with an alternate EoS. However, these changes would not alter the qualitative behaviors we discuss. Consequently, the main conclusions drawn in our study are independent of the specific choice of EoS.

In our study of pulsar spin-down evolution as described by Eq.~\eqref{eq:SD:dimless:slow}, we omit the consideration of magnetic field evolution and concentrate on the novel BNV effects. Nonetheless, extending our study to account for these effects remains straightforward. The theoretical framework surrounding the origin, structure, and evolution of the $H$ field is a active area of research~(refer to~\cite{Igoshev:2021ewx} and the citations therein for an overview). There exists a range of observational evidence underscoring the decay of the magnetic field in neutron stars. For instance, the absence of observed objects with $P \gtrsim 20$ s in Magnetars~\cite{Kaspi:2017fwg}, as well as their potential descendants, the Magnificent Seven (M7)~\cite{Turolla2009}, support this view. Positing that a neutron star core becomes a type I superconductor during its cooling phase~\cite{BAYM1969}, the magnetic field would consequently be repelled from the core due to the Meissner effect~\cite{Meissner}. The evolution of the crustal magnetic field is dictated by a combination of Faraday's law with Ohm's law~\cite{1992ApJ...395..250G}. This results in two components, namely, Ohmic decay and the Hall effect. As the magnetic field undergoes decay, the Hall timescale becomes longer, rendering its effects negligible for a majority of radio pulsars. In this paper, our stance remains neutral regarding the early evolution phases of pulsars. However, for a more complete picture, it would be beneficial to append solutions from models that expound on young pulsar evolution (for $t<0$) to our existing solution, ensuring a matching boundary condition at $t=0$. It is worth noting that even in scenarios where $H$ remains unchanged, the magnetic dipole undergoes alterations due to radius variations, which are brought about by BNV.

To concisely recapitulate, our analysis is founded on these assumptions:
\begin{enumerate}
    \item The rate of BNV is less rapid than the hydrodynamic and chemical equilibration processes in the neutron star, allowing for a quasi-static treatment of the star's response.
    \item  BNV does not alter the original baryon-conserving EoS, thus preserving the unique two-dimensional parameter space derived from it. This assumption is in line with the ``depletion dominance'' scenario explored in Ref.~\cite{universe10020067}, characterized by a more rapid depletion rate of dark particles compared to their production.
    \item Angular momentum loss rates due to BNV ($\dot{J}_{\rm BNV} / J$) are considered negligible relative to the rate of change in the moment of inertia ($\dot{I}_{\rm BNV} / I$), thereby allowing our analysis to remain largely decoupled from the specific particle physics mechanisms driving BNV.
    \item Rotation-induced effects on the structure are considered less significant compared to BNV, quantified by the condition $g(I) < 2 b(I) \delta$.
    \item The magnetic field's evolution does not significantly impact our analysis within the particular time intervals of interest, enabling us to neglect its temporal variation.
\end{enumerate}
Our methodology is primarily designed around the first two premises, ensuring a robust baseline for exploring BNV dynamics. The potential non-applicability of the latter three assumptions is already addressed within our formalism, as detailed in this section. Furthermore, scenarios involving a violation of the last two assumptions have been thoroughly investigated in existing literature. Consequently, our focus is to unveil and scrutinize the novel effects arising specifically from BNV within this established framework. 

With these foundational assumptions clearly established, we turn our attention to the hitherto unexplored BNV effects, with the goal of shedding light on their influence on pulsar physics.

\section{Evolution of Pulsars}
\label{sec:pulsar:evol}
\begin{figure}[t]
    \centering
    \includegraphics[width=0.7\textwidth]{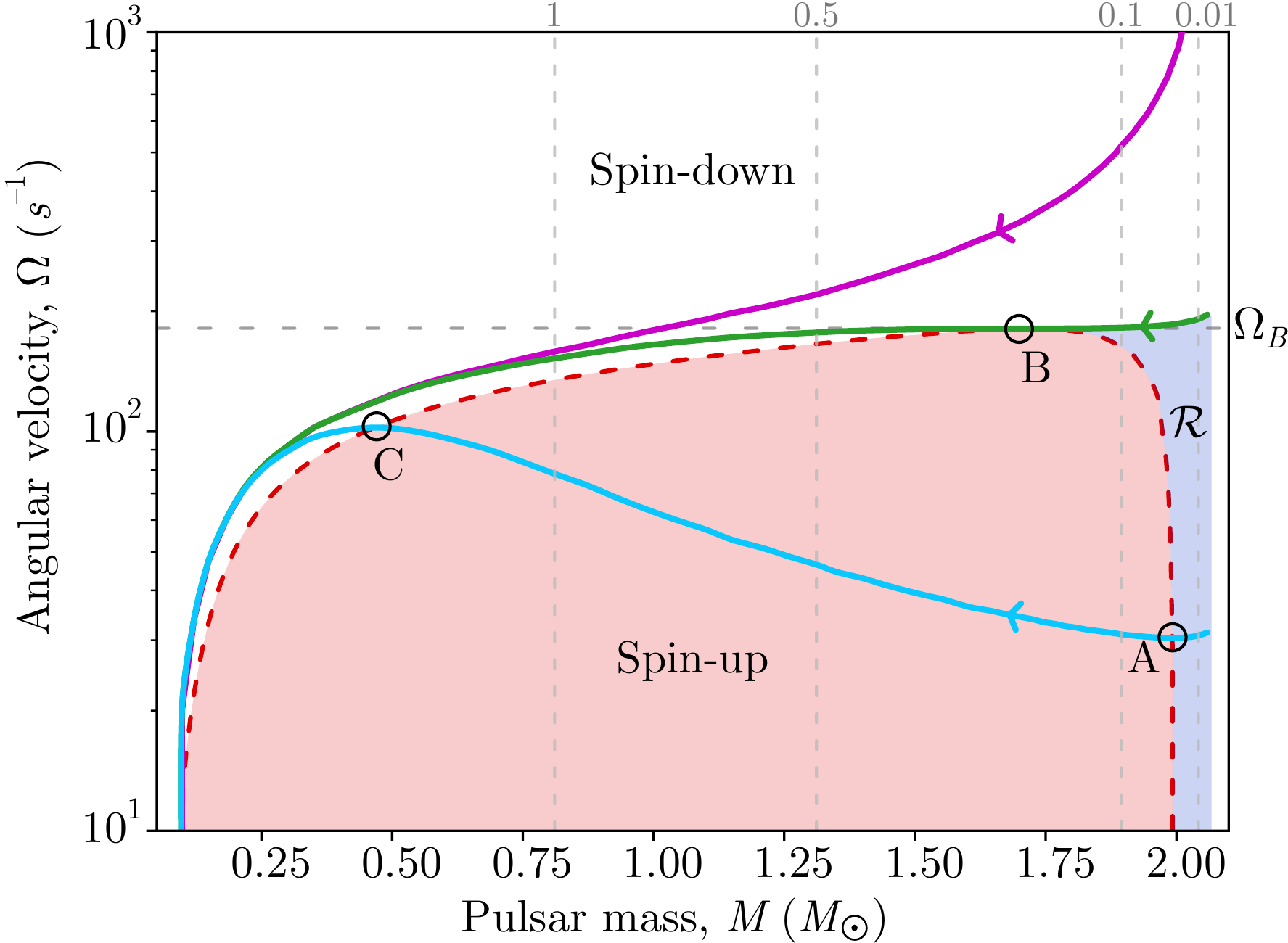}
    \caption{The three distinct types of pulsar evolution trajectories induced by quasi-equilibrium baryon loss. Pulsars within region $\mathcal{R}$ exhibit a transition from spinning down to spinning up. The cyan trajectory illustrates a pulsar beginning in region $\mathcal{R}$ that transitions to spinning up at point A and subsequently reverts to spinning down at point C. Trajectories above the green curve remain in a continuous spin-down phase without any transitional behavior. The temporal evolution is marked by vertical dashed lines at times $t = \{0.01, 0.1, 0.5, 1\}\, \Gamma^{-1}$, where $\Gamma$ represents the rate of baryon loss. 
    }
    \label{fig:omega_m:schem}
\end{figure}

\begin{figure}[t]
    \centering
    \includegraphics[width=0.8\textwidth]{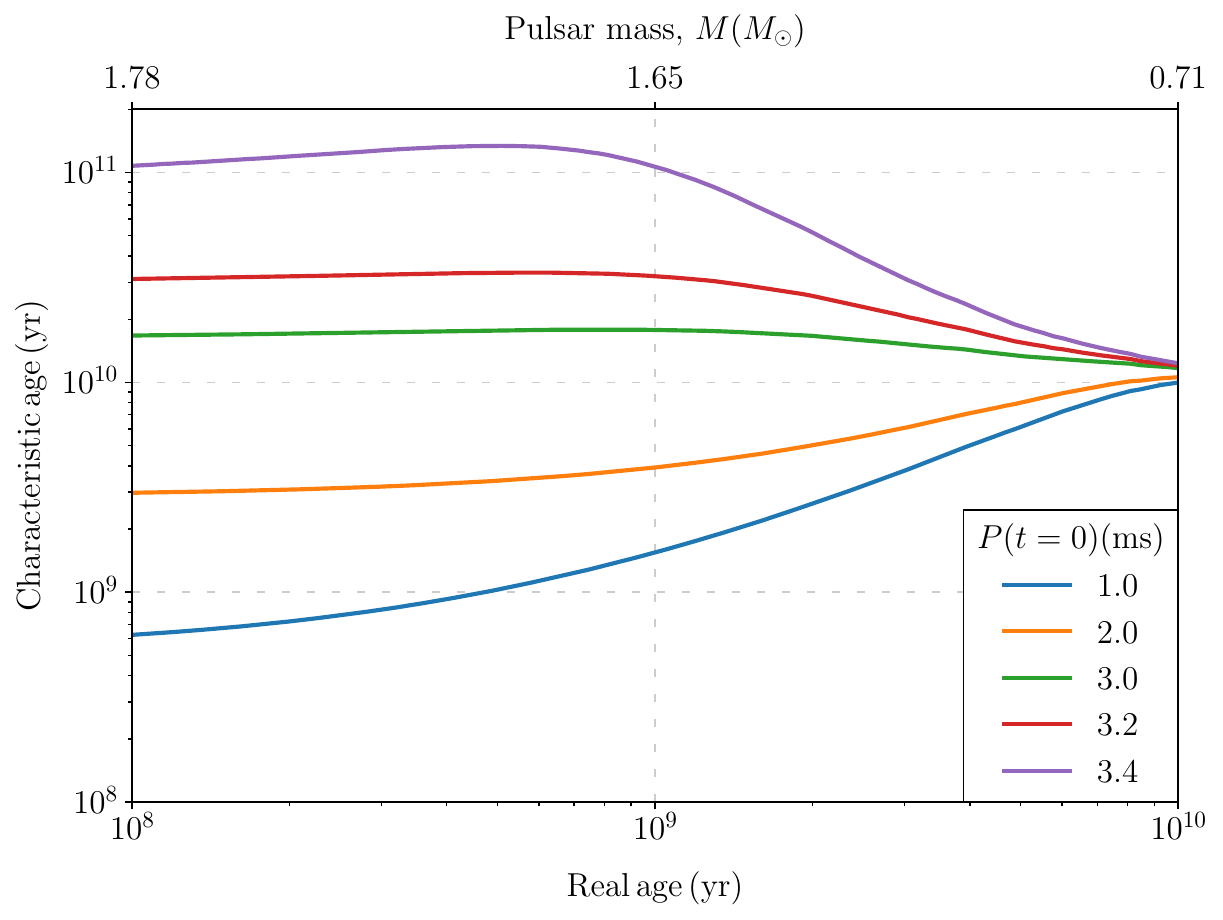}
    \caption{Characteristic ages $P/2\dot{P}$ plotted against the real ages of five pulsars. All pulsars start with an initial mass of $M (t=0) =1.8\, M_{\odot}$. They have different initial periods $P (t=0)$ ranging from $1.0\, {\rm ms}$ to $3.4 \, {\rm ms}$, close to the point $(M, P) = (1.8 M_{\odot}, 3.5\, {\rm ms})$ on the $\dot{\Omega}=0$ curve. We assumed the values $\Gamma = 10^{-10}\, {\rm yr}^{-1}$ and $H=10^8\, {\rm gauss}$.}
    \label{fig:sd_age_vs_t}
\end{figure}

We begin by observing that for $b(I) \delta = 1$, we have $\dot{\Omega}=0$. This sets the boundary between spin-down and spin-up solutions for $M < 2 M_{\odot}$ ($b(I) > 0$) at
\begin{equation}
    \frac{P_{\rm SD-SU}}{{\rm ms}} \approx 3.71\, \left(\frac{H}{10^8\, {\rm gauss}}\right) \left(\frac{R}{12\, {\rm km}}\right)^3 \left(\frac{10^{-10}\, {\rm yr}^{-1}}{b(I)\, \Gamma}\right)^{1/2} \left(\frac{10^{45}\, {\rm g}\, {\rm cm}^2}{I}\right)^{1/2},
    \label{eq:p:trans}
\end{equation}
where solutions spin-up for $P > P_{\rm SD-SU}$ and vice versa. The second derivative of angular frequency at this juncture is
\begin{equation}
    \left[\ddot{\Omega}\right]_{\dot{\Omega}=0} = -b(I) \Gamma^2 \Omega \left[ 1 +  b (\beta(I)) - 6 b(R)\right], \label{eq:omega_ddot_extrem}
\end{equation}
corresponding inversely to the cyan curve in figure~\ref{fig:b-factor}. As illustrated in figure~\ref{fig:b-factor}, we deduce that $[\ddot{\Omega}]_{\dot{\Omega}=0} < 0$ for $M < M_B \sim 1.7\, M_{\odot}$ and is positive otherwise. This implies potential pulsar transitions between spinning-down and spinning-up states. Examining the $(M,\Omega)$ plane (figure~\ref{fig:omega_m:schem}), we note that solutions must exhibit decreasing mass and can only possess turning points at intersections with the $\dot{\Omega}=0$ contour. Hence, a transition from spinning-down to spinning-up (or the inverse) necessitates the pulsar's mass to be greater (or lesser) than $M_B$. Under the presumption of $H$ and $\Gamma$ remaining roughly constant over time, the class of solutions to Eq.~\eqref{eq:SD:dimless:slow} with a fixed ratio $\Gamma/H^2$ during that period are equivalent, barring the time parameter's redefinition. This implies identical pulsar evolution within the same class but at varied speeds. Furthermore, uniqueness of solutions corresponding to various initial conditions ensures that solutions in the same class do not intersect. Pulsars in region $\mathcal{R}$ of figure~\ref{fig:omega_m:schem}, specified by 
\begin{equation}
    \dot{\Omega}<0,\quad M>M_B, \quad {\rm and}\quad \Omega \lesssim \Omega_B\sim 1.8 \times 10^{3}\left(\frac{\Gamma}{10^{-10}\, {\rm yr}^{-1}}\right)^{1/2} \left(\frac{10^{8}\, {\rm gauss}}{H}\right)\, s^{-1}, 
\end{equation}
fulfill the necessary and sufficient conditions for a transition from spinning-down to spinning-up. Using Eq.~\eqref{eq:omega_ddot_extrem}, the timescale for a minor change $\delta\Omega$ around the $\dot{\Omega}=0$ contour is estimated as 
\begin{equation}
    \delta t \approx \left|\frac{2\delta \Omega }{ \ddot{\Omega}}\right|^{1/2}  = \frac{\sqrt{2}\, \Gamma^{-1} }{\sqrt{\left|b(I) \left[ 1 +  b (\beta(I)) - 6 b(R)\right]\right|}}  \sqrt{\frac{|\delta \Omega| }{ \Omega }} \gtrsim 0.31\, \Gamma^{-1} \sqrt{\frac{|\delta \Omega| }{ \Omega }},
\end{equation}
implying that witnessing such a transition within $\delta t \sim 100$ yr (given $\Gamma \sim 10^{-10}\, {\rm yr}^{-1}$) requires the pulsar to be extremely close to the $\dot{\Omega}$ contour with $|\delta \Omega| / \Omega \lesssim 10^{-15}$. The proximity of a pulsar to the $\dot{\Omega} = 0$ contour can lead to a misrepresentation of its age via the characteristic age formula ($P/2\dot{P}$). This discrepancy is portrayed in figure~\ref{fig:sd_age_vs_t} for five pulsars with an initial mass $M (t=0) =1.8\, M_{\odot}$, BNV rate of $\Gamma = 10^{-10}\, {\rm yr}^{-1}$, and a magnetic field $H=10^8\, {\rm gauss}$. The pulsars closer to the $\dot{\Omega} = 0$ contour seem much older based on their characteristic ages. We consider the variations in the braking index for pulsars nearing this contour in section~\ref{sec:bridx}.

All spinning-up pulsars would eventually transition to spinning-down by crossing the $\dot{\Omega}=0$ contour. However, possible early BNV deactivation can occur once a pulsar's core density drops below a certain threshold~\cite{Berryman:2022zic}. After BNV deactivation, the standard spin-down mechanism would take over.

\begin{figure}[t]
    \centering
    \includegraphics[width=0.8\textwidth]{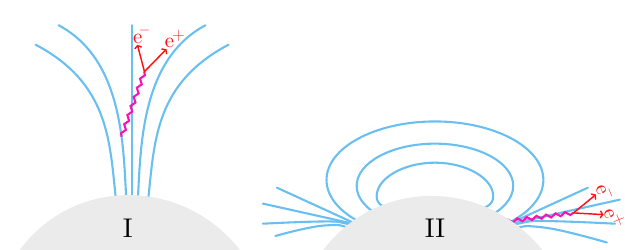}
    \caption{The structure of the magnetic field at the polar cap region of pulsars corresponding to  a pure central dipole (I), and a twisted magnetic field (II)~\cite{1993ApJ...402..264C}. }
    \label{fig:magnetic_structure}
\end{figure}

We now examine the conditions that dictate the transitions between spinning up and spinning down before the pulsar mechanism shuts down~\cite{1971ApJ...164..529S}. The period at which pulsars deactivate is influenced by the neutron star's surface magnetic field structure~\cite{1993ApJ...402..264C}. Two such structures are considered (see figure~\ref{fig:magnetic_structure}), forming the boundaries of the ``death valley'' as outlined in Ref.~\cite{1993ApJ...402..264C}. 
The first structure (I) presents a pure dipole field and has a turnoff period given by
\begin{equation}
    \frac{P_{\rm crit}^{\rm (I)}}{{\rm ms}} \approx 6.1\, \left(\frac{ H_p}{10^{8}\, {\rm gauss}}\right)^{8/15} \, \left(\frac{R}{12\, {\rm km}}\right)^{19/15}.
    \label{eq:p:crit:central}
\end{equation}
While the second structure (II) exhibits a much smaller polar cap area and highly curved field lines. Its turnoff period is
\begin{equation}
    \frac{P_{\rm crit}^{\rm (II)}}{{\rm ms}} \approx 25.4\, \left(\frac{ H_p}{10^{8}\, {\rm gauss}}\right)^{7/13} \, \left(\frac{R}{12\, {\rm km}}\right)^{17/13}.
    \label{eq:p:crit:twisted}
\end{equation}
The spin transitions due to BNV can occur before the pulsar turnoff line if
\begin{equation}
    \frac{\Gamma^{\rm (I)}}{10^{-10}\, {\rm yr}^{-1}} \gtrsim \left(\frac{0.37}{b(I)}\right) \left(\frac{H}{10^8\, {\rm gauss}}\right)^{14/15} \left(\frac{R}{12\, {\rm km}}\right)^{52/15} \left(\frac{10^{45}\, {\rm g}\, {\rm cm}^2}{I}\right),\label{eq:SU_before_death:central}
\end{equation}
in case (I), and 
\begin{equation}
    \frac{\Gamma^{\rm (II)}}{10^{-11}\, {\rm yr}^{-1}} \gtrsim \left(\frac{0.21}{b(I)}\right) \left(\frac{H}{10^8\, {\rm gauss}}\right)^{12/13} \left(\frac{R}{12\, {\rm km}}\right)^{44/13} \left(\frac{10^{45}\, {\rm g}\, {\rm cm}^2}{I}\right),\label{eq:SU_before_death:twisted}
\end{equation}
in case (II).

\begin{figure}[t]
    \centering
    \includegraphics[width=0.7\textwidth]{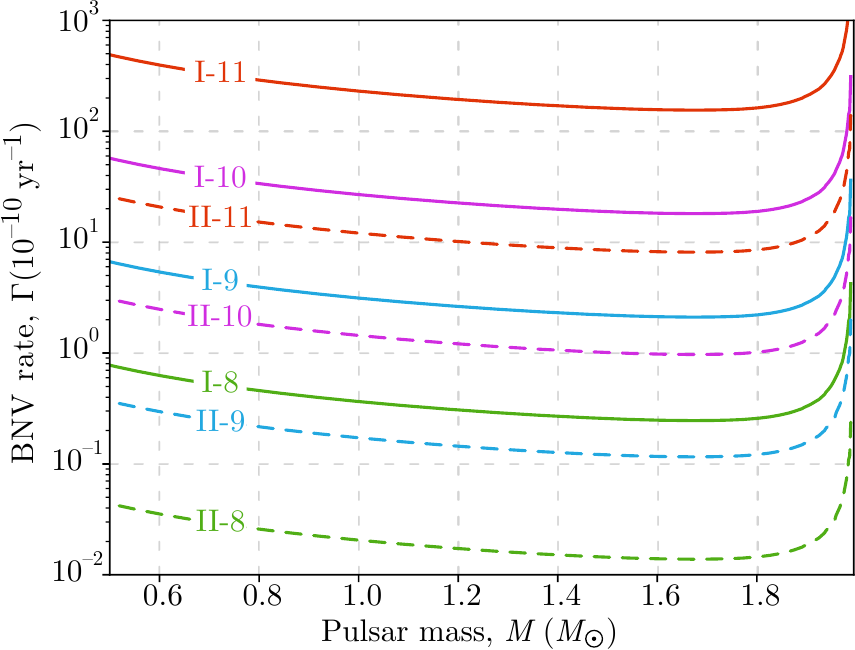}
    \caption{The minimum value of BNV rate ($\Gamma$) in units of $10^{-10}$ per year, needed for the spin-up solutions to exist before the death line. The curve labels correspond to different structure (I: central, II: twisted) and values of the magnetic field ($\log \left( H\, {\rm gauss}\right) = \{ 8, 9, 10, 11\}$). Note that the transition from spinning-down to spinning-up is only possible for $M \gtrsim 1.7\, M_{\odot}$.}
    \label{fig:SU_before_death}
\end{figure}

BNV rates that meet the approximations in these equations could facilitate a revival of dead pulsars or delay their turnoff. Observations of pulsars nearing turnoff, such as PSR B1931+24 which is active only $30\%$ of the time, could lead to insight into the pulsar graveyard region near the turnoff line~\cite{doi:10.1126/science.1125934}. However, pulsars might not exhibit any spin-up due to BNV if its rate is significantly slower than what is described in Eqs.~\eqref{eq:SU_before_death:central} and~\eqref{eq:SU_before_death:twisted}. The minimum BNV rates fulfilling these conditions, as a function of the pulsar's mass and magnetic field, are illustrated in figure~\ref{fig:SU_before_death}.

\begin{figure}[t]
    \centering 
    \includegraphics[width=\textwidth]{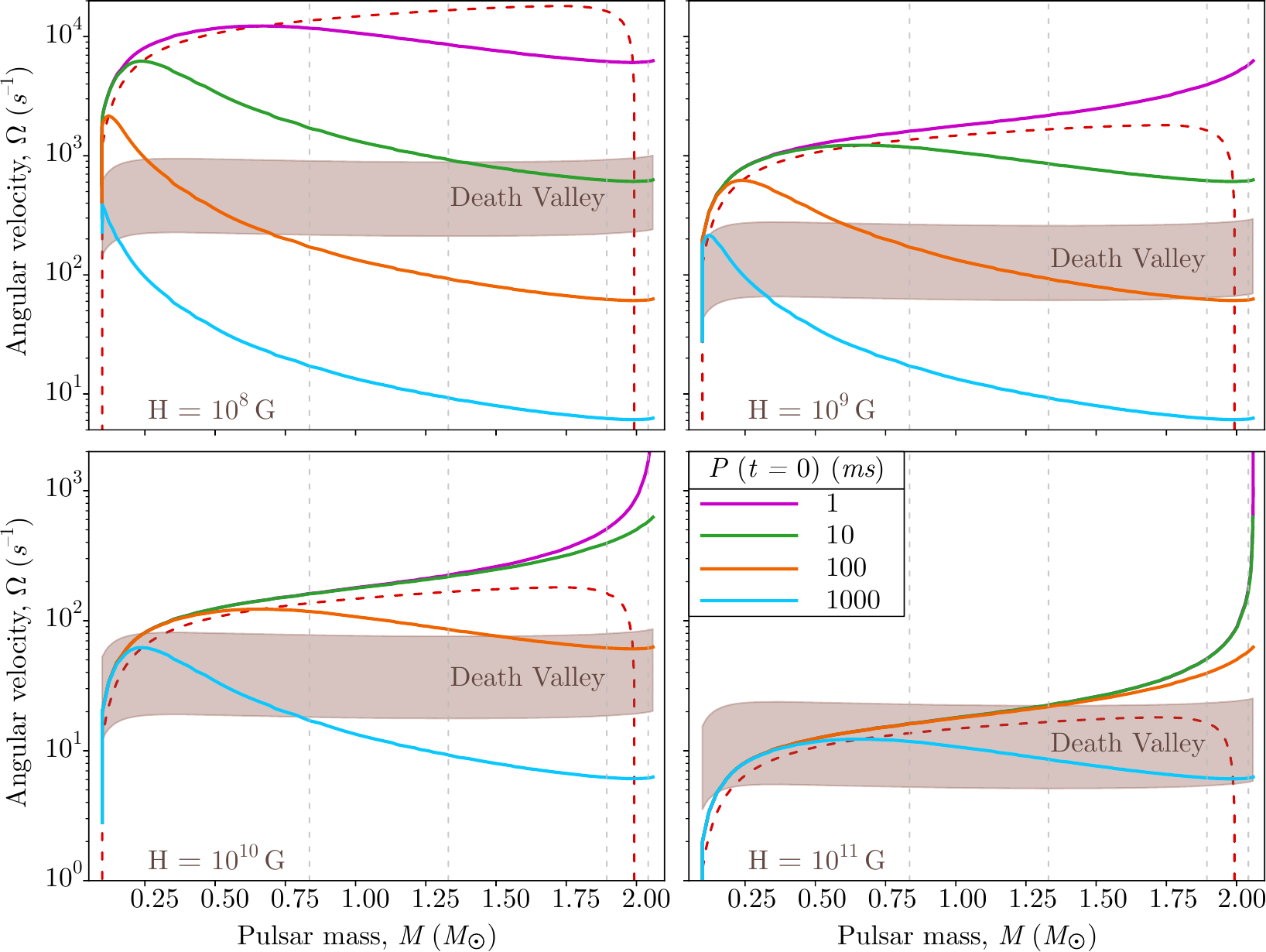}
    \caption{The pulsar evolution trajectories driven by quasi-equilibrium baryon loss at a rate $\Gamma = 10^{-8}\, {\rm yr}^{-1}$, for various magnetic fields ($H$) and initial periods ($P (t=0)$). The brown band shows the death valley region, beneath which pulsar emission ceases. The time epochs $t = \{0.01, 0.1, 0.5, 1\}\, \Gamma^{-1}$ (increasing from right to left) are shown with vertical dashed lines. Note that trajectories with the same value of $\Gamma/H^2$ are equivalent.}
    \label{fig:omega_m:numer}
\end{figure}

In the $\left(M, \Omega\right)$ plane, we plot the evolution of pulsars for a range of magnetic fields and initial periods, as depicted in figure~\ref{fig:omega_m:numer}. The death valleys are represented by a brown band; beneath this, the pulsar emission mechanism ceases. Vertical dashed lines mark time epochs of $t = \{0.01, 0.1, 0.5, 1\}\, \Gamma^{-1}$. It is worth noting that our assumption here is that as pulsars traverse the death valley, their spin-down remains unaffected by the turnoff mechanism. 
However, pulsar wind braking models predict a decreased spin-down rate after a pulsar's ``death.'' An illustrative observational piece supporting this theory is the behavior of PSR B1931+24~\cite{2006Sci...312..549K}, characterized by 
an active pulsation for 5 to 10 days, followed by a quiescent phase lasting about 25 to 35 days, and a slowed spin-down rate during its quiescent phases.


\subsection{Braking Index}
\label{sec:bridx}

In this section, we explore the effects of BNV on the pulsar braking index. Within the framework established by our study, the braking index is given by:
\begin{equation}
\begin{split}
n \equiv \frac{\ddot{\Omega}\Omega}{\dot{\Omega}^2} 
= 3 &+ \frac{\left[b(I) + 6 b(R)\right]\delta - b(I) \delta^2 \left[1 + b(\beta(I)) + b(I) \right]}{\left( 1 - b(I) \delta\right)^2} \\
&- \frac{\delta}{\Gamma \left( 1-b(I) \delta\right)^2} \left[ \frac{2 \dot{H}}{H} + \frac{\kappa_1\, \sin(2\alpha) \, \dot{\alpha}}{\kappa_0 + \kappa_1 \sin^2(\alpha)}\right], 
\label{eq:n:slow}
\end{split}
\end{equation}
in which we neglected the rotation effects by dropping all the $g(\mathcal{O})$ factors, $\dot{\alpha}$ is given by Eq.~\eqref{eq:alpha}, and assumed that BNV rate is constant in time $\dot{\Gamma}_{\rm BNV} \approx 0$. The numerical values for the $b$-factors needed to evaluate $n$ in Eq.~\eqref{eq:n:slow} are plotted in figure~\ref{fig:b-factor}. We also set $\dot{H} = \dot{\alpha} = 0$, as we did in previous sections.

We may observe the deviation in braking index due to BNV in pulsars with $\Gamma \gtrsim \dot{P}/P$, because we expect that a change in the dipole moment, at a rate $\Gamma$, would modify the braking index according to $n_{\rm obs} \sim 3 - \Gamma (P/\dot{P}) $~\cite{1988MNRAS.234P..57B}. We first rewrite the spin-down equation~\eqref{eq:SD:dimless:slow} using the definition of $\delta$ in Eq.~\eqref{eq:defs}
\begin{equation}
    x\equiv - \frac{\dot{\Omega}}{\Omega \Gamma} = \delta^{-1} - b(I),
\end{equation}
and solve for $\delta$, then, plug $\delta = 1/[x + b(I)]$ in the braking index equation~\eqref{eq:n:slow} to derive an equation for $x$:
\begin{equation}
    \left(n-3\right) x^2 - \left[b(I) + 6 b(R)\right] x + b(I) \left[ 1 + b(\beta(I)) - 6 b(R)\right] = 0, \label{eq:x}
\end{equation}
which shows that there is also a nontrivial ($\delta \neq 0$) solution to $n=3$. 

Figure~\ref{fig:n(M,x)} depicts the braking index contours for both spinning-down ($x>0$) and spinning-up ($x<0$) pulsars. These contours are categorized into three sets: (a) $2.8 \leq n \leq 3.61$, (b) $n < 2.8$, which are exclusively observed when $M < 1.72\, M_{\odot}$, and (c) $n > 3.61$, observed solely when $M > 1.72\, M_{\odot}$. Notably, significant deviations from $n=3$ are evident in regions close to $x = 0$ and in the vicinity of the maximum mass.

Figure~\ref{fig:x(M,n)} exhibits contours representing various $x$ values in the $(M, n)$ plane. Regions without solutions are shaded in light gray. For pulsars with a typical mass of $1.4 M_{\odot}$, the feasible range is limited to $n\lesssim 3.35$. On the other hand, pulsars with masses $M \gtrsim 1.72 M_{\odot}$ are restricted to $n \gtrsim 2.8$. It is important to note that alterations to the EoS may shift these mass values, but they will not change the basic shape of these curves.

An alternative expression derived from Eq.~\eqref{eq:x}, defined in terms of $\delta \propto \Gamma / H^2\, \Omega^2$ (see Eq.~\eqref{eq:defs}), is given by:
\begin{equation}
\left\{\left(n-2\right) \left[b(I)\right]^2 + b(I) \left[1 + b(\beta(I))\right] \right\} \delta^2 + \left[\left(5 - 2n\right) b(I) - 6 b(R)\right] \delta + n - 3= 0. 
\end{equation}
This expression is utilized to plot contours of the braking index ($n$) in Figure~\ref{fig:n(M,Omega)} against the pulsar's mass ($M$) and its scaled angular velocity, $\Omega \times H_9/\sqrt{\Gamma_{10}}$. Here, $H_9$ represents $H / 10^{9}\, {\rm gauss}$ and $\Gamma_{10}$ stands for $\Gamma / 10^{-10}\, {\rm yr}^{-1}$. The angular velocities associated with various $(H, \Gamma)$ values are obtained by adjusting $(H_9, \Gamma_{10})$ accordingly. Figure~\ref{fig:n(M,Omega)} displays that pulsars nearing the $\dot{\Omega}=0$ contour (dashed-green curve) exhibit a surge in their braking indices. The rate of change in the braking index of a spinning-down pulsar approaching the $\dot{\Omega}=0$ contour is estimated by:
\begin{equation}
\lim_{\dot{\Omega}\to 0} \dot{n} \approx 2 \left\{ b(I) \left[ 1 +  b (\beta(I)) - 6 b(R)\right]\right\}^2 \Gamma^4 \left(\frac{P}{\dot{P}}\right)^3,
\end{equation}
in which the numerical factor behind $\Gamma^4 (P/\dot{P})^3$ exceeds $0.1$ for pulsars heavier than $M_B \sim 1.7 M_{\odot}$ and is below $\sim 30$ for pulsars lighter than $2 M_{\odot}$. The timescale for a change $\delta n$ in the braking index can then be approximated by:
\begin{equation}
5.5 \times 10^{-4} \left(\frac{\delta n}{\Gamma^4}\right) \left(\frac{\dot{P}}{P}\right)^3 \leq \delta t <  50 \left(\frac{\delta n}{\Gamma^4}\right) \left(\frac{\dot{P}}{P}\right)^3 \qquad\qquad M_B < M \leq 2 M_{\odot}.
\end{equation}
Consequently, the braking index of a spinning-down pulsar nearing the $\dot{\Omega}=0$ contour will increase by more than one unit ($\delta n > 1$) in a decade, provided $\dot{P}/P < f \times 10^{-20} \Gamma_{10}$, where $f \in [0.09, 4]$ depending on the pulsar's mass.
 
A diverse set of braking index values can be generated within our framework, offering a plausible explanation for the observed braking indices of young pulsars. Consider the Crab pulsar (B0531+21) with parameters $P \approx 33\, {\rm ms}$, $\dot{P} = 4 \times 10^{-14}$, and a braking index of $n=2.5$~\cite{10.1093/mnras/stu2118}. This particular braking index requires a BNV rate of approximately $\Gamma \approx 10^{-5}$ per year for its explanation. If we presume that BNV results from specific particle physics interactions, universally affecting pulsars of similar (and heavier) masses, it follows that pulsars with $M\geq M_{\rm Crab}$, and older than $\Gamma^{-1} \sim 10^5$ years should not exist, as they would have plausibly depleted their mass due to these interactions. However, we cannot dismiss the possibility of a non-universal BNV mechanism, for instance, one reliant on the core temperature of neutron stars, thus potentially active only in younger (hotter) pulsars.

Numerous alternative explanations exist for the observed low braking indices in young pulsars~\cite{Zhang:2022dso}. Effects due to changes in the moment of inertia of pulsars with superfluid cores~\cite{ho2012rotational}, magnetic field evolution~\cite{10.1093/mnras/204.4.1025, 1984Ap&SS.102..301B, Tauris:2001cy, PhysRevD.94.063012, 10.1093/mnras/stu2140,doi:10.1126/science.1243254}, modifications to modeling the magnetosphere~\cite{10.1093/mnras/288.4.1049, Contopoulos:2005rs}, torque from particle wind flows~\cite{1969ApJ...157..869G, 1969ApJ...158..727M, MICHEL1969pulsaremission}, and internal damping of oscillations in wobbling pulsars~\cite{Araujo:2023unj} are among the notable examples. 

However, these considerations do not necessarily eliminate the prospect of a universal BNV occurring at a rate slower than the inverse average age of the oldest known (massive) pulsars. Notably, the effects of BNV might become apparent in older pulsars, indicated by smaller $\dot{P}/P$ values, provided they possess sufficient mass to accommodate BNV processes. In the ensuing section, we delve into the potential ramifications of these subtle BNV effects for precision pulsar timing.

\begin{figure}[t]
    \centering
    \includegraphics[width=\textwidth]{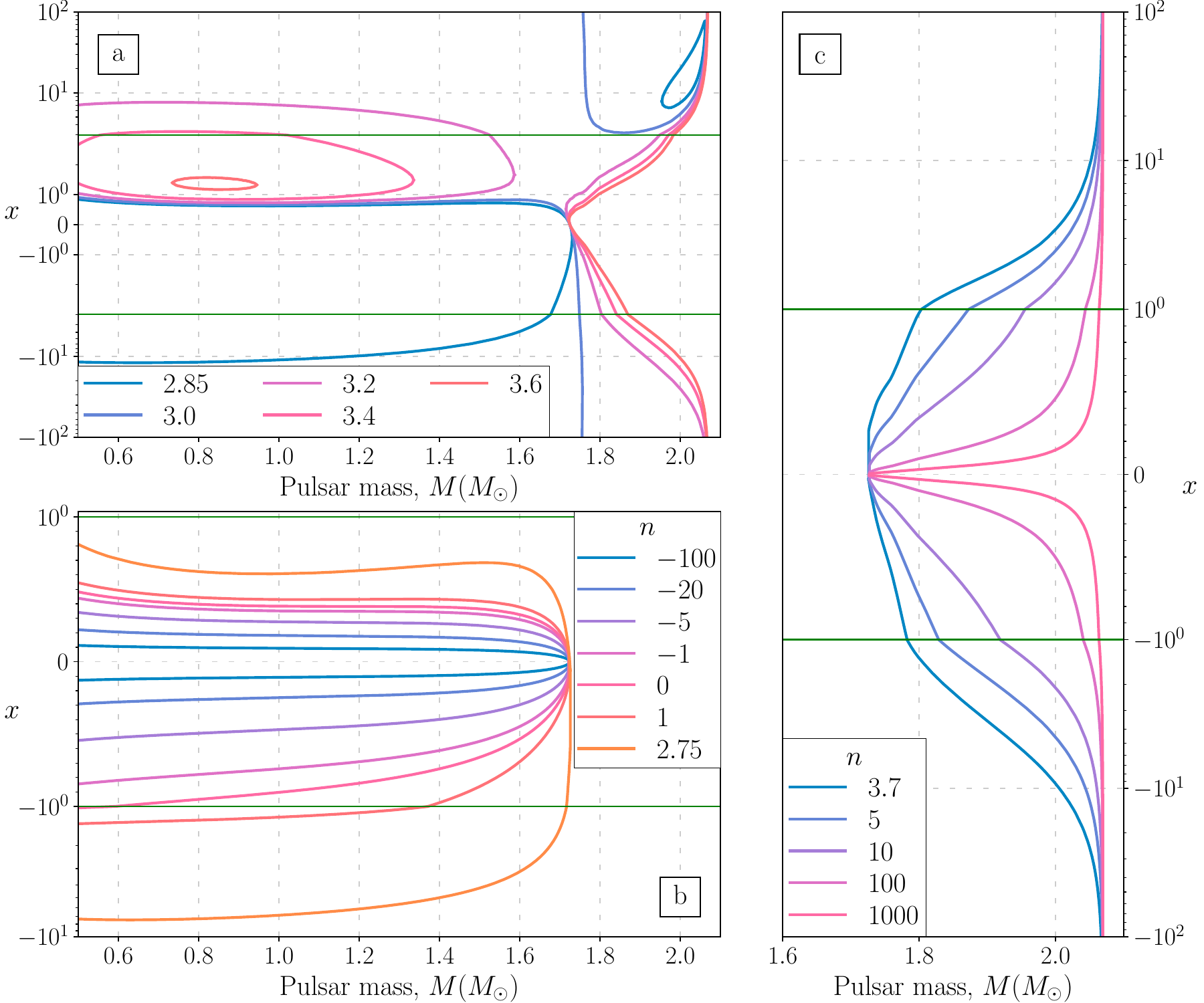}
    \caption{
    Contours of the braking index ($n$) as a function of the pulsar's mass ($M$) and dimensionless spin-down rate $x \equiv (\dot{P}/P) / \Gamma$, using CMF-1~\cite{Dexheimer:2008ax} as the EoS. Symmetrical logarithmic scaling is applied to the y-axis, with a linear range of $[-3, 3]$ in panel (a) and $[-1, 1]$ in panels (b) and (c). The transitions to logarithmic scaling are marked by solid green horizontal lines.
    }
    \label{fig:n(M,x)}
\end{figure}
\begin{figure}[t]
    \centering
    \includegraphics[width=0.85\textwidth]{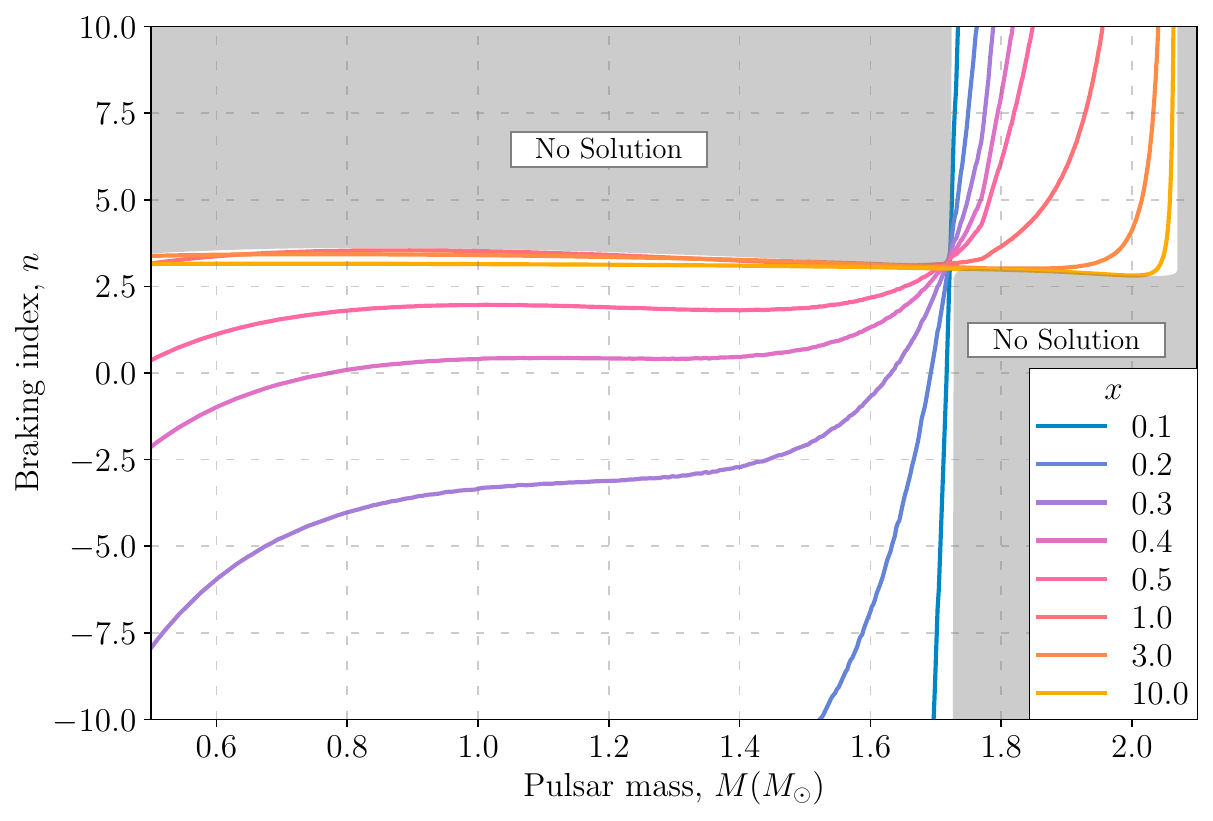}
    \caption{Contours of $x \equiv (\dot{P}/P) / \Gamma$ as a function of the pulsar's mass ($M$) and braking index ($n$) for a choice of CMF-1~\cite{Dexheimer:2008ax} as EoS. There are no solutions in the gray areas.}
    \label{fig:x(M,n)}
\end{figure}

\begin{figure}[t]
    \centering
    \includegraphics[width=0.85\textwidth]{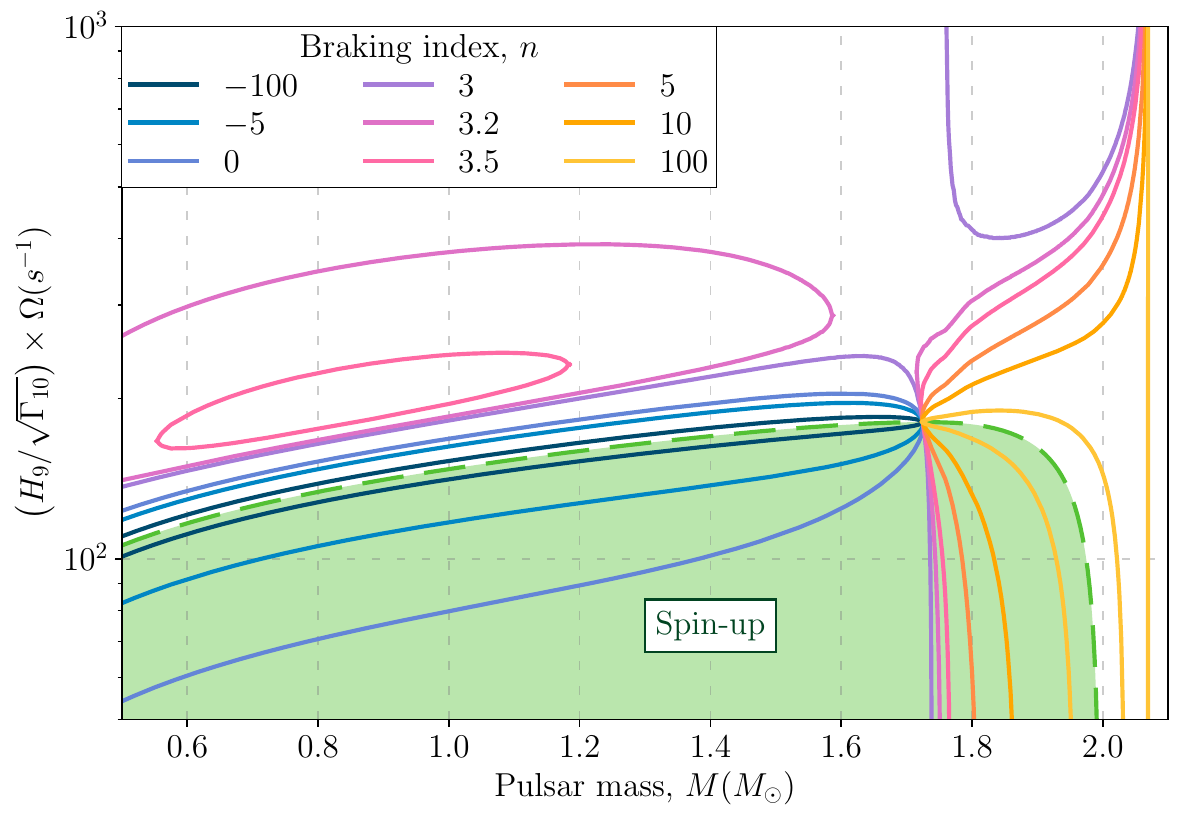}
    \caption{Contours of braking index ($n$) a function of a pulsar's mass ($M$) and its angular velocity ($\Omega$). The Pulsars below the dashed-green curve ($\dot{\Omega}=0$) spin-up and those above it spin-down. The y-axis is scaled by the ratio $H_9/\sqrt{\Gamma_{10}}$, in which $H_9 \equiv H / 10^{9}\, {\rm gauss}$, and $\Gamma_{10} \equiv \Gamma / 10^{-10}\, {\rm yr}^{-1}$.}
    \label{fig:n(M,Omega)}
\end{figure}

\section{Observation Prospects}\label{sec:obsprosp}

We now delve into exploring BNV signatures influencing the second derivative of pulsar spin frequency ($\ddotnu$). Measurements of $\ddotnu$ face challenges due to stochastic irregularities in rotation occurring over months to years (timing noise), and abrupt changes in $\nu$ (glitches) which are followed by gradual relaxations. These effects contribute to large and time-dependent values of $n$ (positive and negative)~\cite{Lower:2021rdo}. 

Post-glitch relaxation from previously unobserved glitches is suggested to introduce a bias towards higher positive $\ddotnu$ in relatively young pulsars~\cite{Johnston:1999ka}. Consequently, only observed values of $\nobs$ that approximate the standard value of $3$ are considered to represent the intrinsic braking indices of pulsars, such as $\sim 0.03$ for \psrB~\cite{Marshall:2016kbp} and around $\sim 3.15$ for \psrJ~\cite{Archibald:2016hxz}. Larger glitches are typically detected in pulsars with characteristic ages under $10^5$ years. A statistical examination of pulsar glitch occurrences~\cite{1994MNRAS.269..849A} suggests that the number of glitches in a sample of pulsars is related to each pulsar's $\dot{P}/P$, indicating less glitch activity in older pulsars~\cite{Zhou:2022cyp}.

The origin of timing noise in pulsars can stem from intrinsic and extrinsic sources. Intrinsic factors may arise from fluctuations in the internal structure of the pulsar~\cite{Melatos:2013rca,Shannon_2010, 1990MNRAS.246..364J} or variations in the magnetospheric torque~\cite{1987ApJ...321..805C,2006Sci...312..549K,2008ApJ...682.1152C,doi:10.1126/science.1186683}. Meanwhile, extrinsic noise can be attributed to measurement errors and the effects of the interstellar medium (ISM)~\cite{1984Natur.307..527A, 1984JApA....5..369B, 1990ApJ...364..123F, 1990ARA&A..28..561R}, which tend to diminish at higher observation frequencies~\cite{Stinebring:2013yza}. Notably, observations by the Neutron Star Interior Composition Explorer (NICER) in the X-ray band offer an opportunity for extended high-precision pulsar timing, mitigating extrinsic noise originating from the ISM~\cite{Hazboun:2021two}. 

The timing noise of millisecond pulsars (MSPs) induced over decades of observation can be lower than $200$ ns level~\cite{Shannon_2010}. Precision timing of MSPs has been utilized for detecting stochastic GW background~\cite{1979ApJ...234.1100D, 1983ApJ...265L..39H}. This involves assessing angular correlations and slight deviations in the timing residuals across a network of stable MSPs forming a Pulsar Timing Array (PTA). Several PTAs have reported significant evidence of a GW background~\cite{Agazie_2023, EPTA:2023fyk, Reardon_2023, Xu_2023}. While distinguishing stochastic GW signals from noise presents challenges due to their similar characteristics, the constant nature of BNV effects over extended periods may allow their differentiation from stochastic variable noise~\cite{1985ApJS...59..343C}. By effectively addressing extrinsic sources of noise, older MSPs exhibiting high-precision timing stand as promising candidates for exploring BNV effects. We evaluate the necessary timing precision required for this endeavor.

In the absence of major glitches or substantial timing noise during the observation period ($T$), the braking index, $n$, remains fairly constant, and its value can be derived without a direct $\ddot{\nu}$ measurement~\cite{Johnston:1999ka}:
\begin{equation}
    n = 1 + \frac{\nu_1 \dot{\nu}_2 - \nu_2 \dot{\nu}_1}{\dot{\nu}_1 \dot{\nu}_2\, T},
\end{equation}
in which indices $1$ and $2$ correspond to the start and end of the observation period.  Assuming the errors in $\dot{\nu}$ measurements to be comparable, i.e., $\sigma(\dot{\nu}_1) \approx \sigma(\dot{\nu}_2) \equiv \sigma(\dot{\nu})$, the anticipated absolute error in $n$ is $\sigma(n) \approx \nu\, \sigma(\dot{\nu}) / (T \dot{\nu}^2)$. An examination of figure~\ref{fig:n(M,x)} and figure~\ref{fig:x(M,n)} reveals that if $x \lesssim 1$, a departure (from $n=3$) of order unity due to BNV in braking index may be seen. In this case we can get a lower bound on $\sigma(n)$:
\begin{equation}
    \sigma(n) \approx \frac{\nu\, \sigma(\dot{\nu})}{T \dot{\nu}^2} \gtrsim 0.5\, \left(\frac{20\, {\rm yr}}{T}\right) \left(\frac{\sigma(\dot{\nu}) / |\dot{\nu}|}{10^{-9}}\right)  \left(\frac{10^{-10}\, {\rm yr}^{-1}}{\Gamma}\right).
\end{equation}
Consequently, to discern changes of order unity in $n$ due to BNV at a rate $\Gamma \sim 10^{-10}\, {\rm yr}^{-1}$, within measurements taken approximately $T \approx 20$ years apart, the relative error in $\dot{\nu}$ should satisfy $|\sigma(\dot{\nu}) / \dot{\nu}| \lesssim 10^{-9}$.

\begin{figure}[t]
    \centering
    \includegraphics[width=0.8\textwidth]{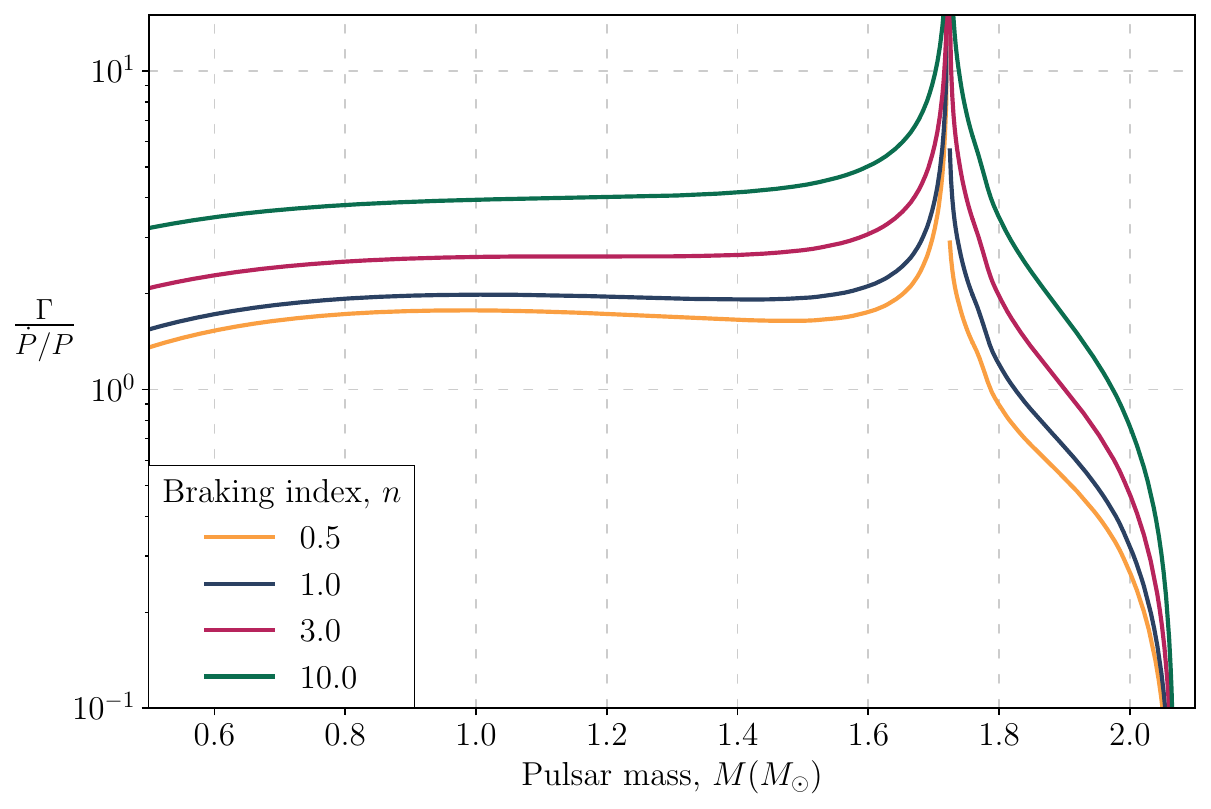}
    \caption{The minimum values of the ratio ${\Gamma}/ {(\dot{P}/P)}$ satisfying Eq.~\eqref{eq:gamma_min:sensitive} as a function of a pulsar's mass ($M$), and for a range of selected braking indices ($n \in \{0.5, 1, 3, 10\}$). Observations on the verge of resolving $\ddot{\nu}$ may be sensitive to BNV rates that fall above these curves.}
    \label{fig:gamma_min:sensitive}
\end{figure}

We write an expression for the BNV contribution to the second derivative of frequency ($\ddot{\nu}_{\rm BNV}$) by combining Eq.~\eqref{eq:SD:dimless:slow} with its second derivative written in terms of frequency ($\nu$) to remove the magnetic field and find 
\begin{equation}
    \ddot{\nu}_{\rm BNV} \equiv \ddot{\nu} - 3\left(\frac{\dot{\nu}^2}{\nu}\right)  = - c_1 \Gamma \dot{\nu} - c_2 \Gamma^2 \nu, \label{eq:nuddot}
\end{equation}
in which the numerical coefficients $c_1 \equiv b(I) + 6 b(R)$ and $c_2 \equiv  b(I) \left[1 + b(\beta(I)) - 6 b(R)\right] $ are plotted in figure~\ref{fig:b-factor} as a function of mass. The condition $|\ddot{\nu}_{\rm BNV}| > \sigma(\ddot{\nu})$ for spinning-down pulsars implies
\begin{align}
    \frac{|c_1|}{2|c_2|} \left[ \pm 1 + \sqrt{1 + \frac{4 |c_2|}{c_1^2} \left(\frac{\sigma(\ddot{\nu})}{\dot{\nu}^2 / \nu}\right) } \right] <& \frac{\Gamma}{\dot{P}/P},
\end{align}
in which we pick the minus sign for $1.72\, M_{\odot} \lesssim M \lesssim 1.76\, M_{\odot}$ and the plus sign otherwise. There is also a range of solutions with smaller values of $\Gamma$ for both $M \lesssim 1.72\, M_{\odot}$ and $1.76\, M_{\odot} \lesssim M$
\begin{align}
        \frac{|c_1|}{2|c_2|} \left[ 1 - \sqrt{1 - \frac{4 |c_2|}{c_1^2} \left(\frac{\sigma(\ddot{\nu})}{\dot{\nu}^2 / \nu}\right) } \right]  <& \frac{\Gamma}{\dot{P}/P} < \frac{|c_1|}{2|c_2|} \left[ 1 + \sqrt{1 - \frac{4 |c_2|}{c_1^2} \left(\frac{\sigma(\ddot{\nu})}{\dot{\nu}^2 / \nu}\right) } \right] 
\end{align}
which only exist if $n \times (\sigma(\ddot{\nu})/ \ddot{\nu}) < c_1^2 / 4 |c_2|$. The values for $c_1^2 / 4 |c_2|$ are plotted using a cyan curve in Fig~\ref{fig:b-factor}. Notably, this range of solutions does not exist for observations of pulsars with $n>0.6$ that are on the verge of resolving $\ddot{\nu}$, where $\ddot{\nu} \sim \sigma(\ddot{\nu})$. Observations of this kind may exhibit sensitivity to BNV effects if they satisfy:
\begin{align}
    \frac{|c_1|}{2|c_2|} \left[ \pm 1 + \sqrt{1 + \frac{12 |c_2|}{c_1^2} \left(\frac{n}{3}\right) } \right] \lesssim& \frac{\Gamma}{\dot{P}/P}. \label{eq:gamma_min:sensitive}
\end{align}
The BNV rates satisfying this condition are illustrated in Figure~\ref{fig:gamma_min:sensitive}.

\begin{figure}[t]
    \centering
    \includegraphics[width=0.85\textwidth]{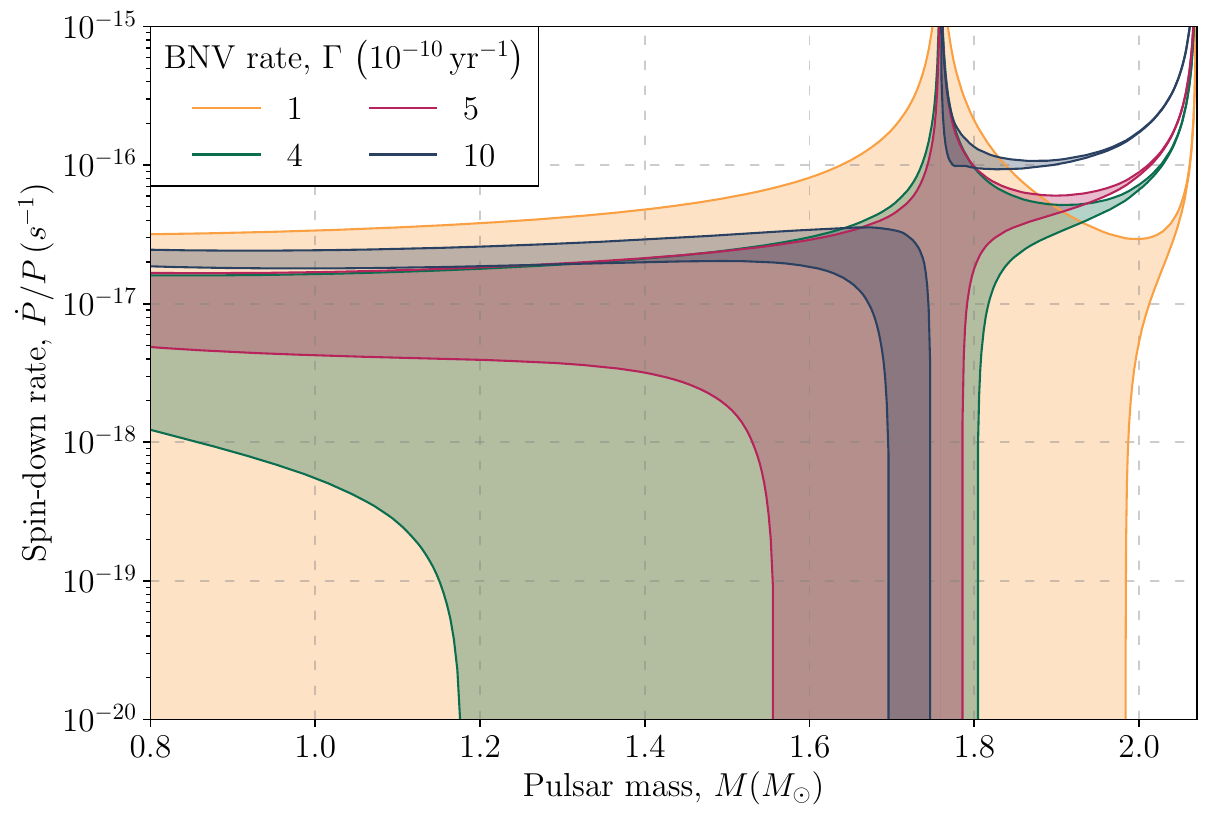}
    \caption{The shaded regions correspond to pairs of $(M, \dot{P}/P)$ values for which BNV effects \emph{cannot} be resolved, i.e., $\ddot{\nu}_{\rm BNV} < \sigma(\ddot{\nu})$. Measurements of $\ddot{\nu}$ for pulsars outside the shaded regions may be sensitive to BNV. Here $\sigma(\ddot{\nu})$ is given by Eq.~\eqref{eq:nuddot_err}, in which we assumed observation parameters $T = 50\, {\rm yr}$, $\sigma_{\rm rms} = 100\, {\rm ns}$, and $\Delta t = 7\, {\rm days}$. }
    \label{fig:nuddot:resolve}
\end{figure}

We now compare $\ddot{\nu}_{\rm BNV}$ in Eq.~\eqref{eq:nuddot} to the expected uncertainty in $\ddot{\nu}$ measurements. We assume that the time of arrival (ToA) measurements of a pulsar have equal, Gaussian root-mean-square (rms) uncertainties of $\sigma_{\rm rms}$. Furthermore, these measurements are equally distributed at a cadence $\Delta t$ over a time span $T$.
The expected uncertainty $\sigma(\ddot{\nu})$ is then approximated by~\cite{Liu:2018lmk}
\begin{equation}
    \sigma(\ddot{\nu}) = 5.9\times 10^{-32}\, {s}^{-3} \left(\frac{{\rm ms}}{P}\right) \left(\frac{T}{50\, {\rm yr}}\right)^{-7/2} \left(\frac{\sigma_{\rm rms}}{100\, {\rm ns}}\right)^{1/2} \left(\frac{\Delta t}{\rm d}\right)^{1/2}. \label{eq:nuddot_err}
\end{equation}
Comparing $\sigma(\ddot{\nu})$ with $\ddot{\nu}_{\rm BNV}$, we infer that a 50-year observation of a pulsar, at a weekly cadence with $\sigma_{\rm rms} = 100\, {\rm ns}$, could be sensitive to BNV rates faster than $\Gamma > 10^{-10}\, {\rm yr}^{-1}$, depending on the mass ($M$) and spin-down rate ($\dot{P}/P$) of the pulsar. The regions in the $(M, \dot{P}/P)$ plane where BNV effects are \emph{indistinguishable} from noise, i.e., $\ddot{\nu}_{\rm BNV} < \sigma(\ddot{\nu})$, are depicted in figure~\ref{fig:nuddot:resolve}. As $\Gamma$ increases, these regions shrink, disappearing for $\Gamma \gtrsim 5\times 10^{-9}$. Thus, an observation with our chosen parameters could be sensitive to BNV rates $\Gamma \gtrsim 5\times 10^{-9}$, regardless of the pulsar's mass ($M$) or spin-down rate ($\dot{P}/P$).

Thus far, our discussion has primarily centered on the impact of BNV effects on the intrinsic values of pulsar spin parameters, which we will denote with an ``$i$'' subscript henceforth. We have set aside extrinsic factors that might influence the observed values. For instance, the acceleration of a pulsar within globular clusters can induce observable levels in $\ddot{\nu}$~\cite{10.1093/mnras/225.1.51P}. In essence, separating dynamical effects from intrinsic contributions like BNV could be a complicated process, especially when combined with pulsar timing precision challenges. The observed value for $\ddot{\nu}$ can be broken down as follows
\begin{equation}
\begin{split}
    \ddot{\nu} \approx& \ddot{\nu}_i - 2 \frac{v_{\perp}^2 }{c\, d} \dot{\nu}_i - 2 \frac{a_{\parallel}}{c} \dot{\nu}_i + 3  \frac{v_{\parallel} v_{\perp}^2}{c\, d^2} \nu_i - 3 \frac{\textbf{v}_{\perp}\cdot \textbf{a}_{\perp}} {c\, d}  \nu_i  - (m v_{\parallel} + b) \nu_i\\
            \equiv& \ddot{\nu}_i + \ddot{\nu}_{\rm Shk} + \ddot{\nu}_{\rm acc} + \ddot{\nu}_{\parallel} + \ddot{\nu}_{\perp} + \ddot{\nu}_{\rm jerk},
            \label{eq:nu_ddot:obs}
\end{split}   
\end{equation}
in which $c$ is the speed of light. Here, $\dot{\nu}_i$ and $\ddot{\nu}_i$ represent the intrinsic first and second derivatives of frequency, respectively, inclusive of $\ddot{\nu}_{\rm BNV}$. The terms $v_{\perp}$ and $v_{\parallel}$ signify the transverse and radial velocities of the pulsar relative to the solar system barycenter (SSB), with $v_{\perp} = \mu d$, where $\mu$ denotes its proper motion and $d$ is the distance to the pulsar. Similarly, $a_{\perp}$ and $a_{\parallel}$ represent the transverse and radial accelerations of the pulsar, respectively. The slope ($m$) and intercept ($b$) parameters can be found in table 1 of Ref.~\cite{Liu:2018lmk}.
We now examine the conditions required for each term to contribute less significantly (in magnitude) than BNV for a BNV-dominated pulsar, described as $\dot{P}_i/P_i \sim b(I) \Gamma$.

The different terms and their effects on pulsar parameters can be enumerated as follows:

\begin{enumerate}
    \item \textbf{Shklovskii ($\ddot{\nu}_{\rm shk}$)}: It originates from the Shklovskii effect~\cite{1970SvA....13..562S}, and would be subdominant to BNV under the following condition:
    \begin{equation}
        \frac{\dot{P}_i}{P_i} \gtrsim 5 \times 10^{-21} \, \text{s}^{-1} \, \left( \frac{d}{\text{kpc}} \right) \left(\frac{\mu}{\text{mas} \, \text{yr}^{-1}}\right)^2 \, c_3, \label{eq:bnv_vs_Shk}
    \end{equation}
    in which $c_{3} \equiv |c_1/b_I - c_2/b_I^2|^{-1}$ is shown in figure~\ref{fig:b-factor}.

    \item \textbf{Acceleration ($\ddot{\nu}_{\rm acc}$)}: It becomes less dominant than BNV when:
    \begin{equation}
        \frac{\dot{P}_i}{P_i} \gtrsim 7 \times 10^{-20} \, \text{s}^{-1} \left( \frac{a_{\parallel}}{10^{-14} \, \text{km/s}^2} \right) \, c_3.
    \end{equation}

    \item \textbf{Radial ($\ddot{\nu}_{\parallel}$)}: It is independent of distance when expressed in terms of proper motion ($\mu$), and would be subdominant to BNV under:
    \begin{equation}
        \frac{\dot{P}_i}{P_i} \gtrsim 3 \times 10^{-18} \, \text{s}^{-1} \left( \frac{v_{\parallel}}{50 \, \text{km/s}} \right)^{1/2} \left(\frac{\mu}{\text{mas} \, \text{yr}^{-1}}\right) \, \sqrt{c_3}.
    \end{equation}

    \item \textbf{Transverse ($\ddot{\nu}_{\perp}$)}: It would be less significant than BNV if:
    \begin{equation}
        \frac{\dot{P}_i}{P_i}  \gtrsim 4 \times 10^{-18} \, \text{s}^{-1} \left( \frac{|\hat{\textbf{v}}_{\perp}\cdot \textbf{a}_{\perp}|}{10^{-14} \, \text{km/s}^2} \right)^{1/2} \left(\frac{\mu}{\text{mas} \, \text{yr}^{-1}}\right)^{1/2} \, \sqrt{c_3}.
    \end{equation}

    \item \textbf{Jerk ($\ddot{\nu}_{\rm jerk}$)}: It remains subdominant when:
    \begin{equation}
        \frac{\dot{P}_i}{P_i}   \gtrsim 10^{-17} \, \text{s}^{-1} \left| \left(\frac{m}{10^{-36} \, \text{km}^{-1} \text{s}^{-1}}\right) \left(\frac{v_{\parallel}}{100 \, \text{km/s}}\right) + \frac{b}{10^{-34} \, \text{s}^{-2}} \right|^{1/2} \, \sqrt{c_3}. \label{eq:bnv_vs_jerk}
    \end{equation}
\end{enumerate}

Note that the above relations are approximate. They hold primarily when BNV is the dominant factor affecting pulsar spin-down, i.e., $\dot{P}_i/P_i\approx b(I) \Gamma$. For a detailed analysis, it is recommended to contrast the measurement uncertainty $\sigma(\ddot{\nu})$ with $\ddot{\nu}_{\rm BNV}$ and evaluate the numerical values of the extrinsic contributions for a specific pulsar.

Consider the binary MSP J1909$-$3744, the properties of which are detailed in table~\ref{tab:psr}. Its narrow pulse profile makes it an excellent candidate for precision timing experiments for detecting nanohertz-frequency gravitational waves~\cite{2013Sci...342..334S, 2015MNRAS.453.2576L, 2018ApJ...859...47A, Agazie_2023}. Observations spanning over a period of 15 years, at a cadence $\Delta t \approx 6.3$ days, using the Nan\c{c}ay Radio Telescope have yielded results showing a timing residual root mean square (rms) of $\sigma_{\rm rms} = 103$ ns~\cite{Liu:2020hkx}. From these observations, we infer that the minimum BNV rate exceeding the measurement error in $\ddot{\nu}$ is $\Gamma \gtrsim 4\times 10^{-9}\, {\rm yr}^{-1}$. We identify the effects due to radial velocity ($v_{\parallel}$) as the dominant contribution to $\ddot{\nu}$ in table~\ref{tab:psr}.  Notably, if future observations produce a value $\ddot{\nu} \lesssim - 2.3\times 10^{-29}\, s^{-3}$, this would create a significant $5\sigma$ discrepancy with the null hypothesis (assuming no BNV) given the average measured radial velocity of $v_{\parallel} = -73 \pm 30\, ({\rm km}/{\rm s})$~\cite{handle:20.500.11811/5753}. Such a tension can be alleviated by adding contributions from BNV at a rate of $\Gamma \sim 8 \times 10^{-9}\, {\rm yr}^{-1}$. Conversely, if an expected value from the null hypothesis, $\ddot{\nu} = -8 \times 10^{-30}\, s^{-3}$, is observed, it would establish an upper bound of $\Gamma < 5 \times 10^{-9}\, {\rm yr}^{-1}$ with a $95\%$ confidence level.

\begin{table}[t]
    \centering
    \def\arraystretch{1.45}
    \begin{tabular}{c|c|c|}
    \cline{2-3} 
        & Name & J1909$-$3744  \\
        \cline{2-3}\cline{2-3}
        & $M_p \, (M_{\odot})$ & $1.492(14)$\\ \hline
        %
        %
        \multicolumn{1}{|c|}{\multirow{2.2}{*}{\rotatebox[origin=c]{90}{ \footnotesize Observed}}}
        & $\nu\, ({\rm Hz})$ & $339.315687218483(1)$ \\\cline{2-3}
        \multicolumn{1}{|l|}{}
        & $\dot{\nu}\, ({s}^{-2})$ & $-1.614795(7)\times 10^{-15}$ \\\hline
        \multicolumn{1}{|c|}{\multirow{2}{*}{\rotatebox[origin=c]{90}{\footnotesize Intrinsic}}}
         & $\dot{\nu}_i\, ({s}^{-2})$ & $-3.11 \times 10^{-16}$ \\\cline{2-3}
        \multicolumn{1}{|l|}{} & $\ddot{\nu}_{i}\, (s^{-3})$ & $8.5 \times 10^{-34}$ \\\hline
        \multicolumn{1}{|c|}{\multirow{5}{*}{\rotatebox[origin=c]{90}{\footnotesize Extrinsic}}}
        & $\ddot{\nu}_{\rm Shk}\, ({s}^{-3})$ & $2.4 \times 10^{-33}$ \\\cline{2-3}
        \multicolumn{1}{|l|}{}  & $\ddot{\nu}_{\rm acc}\, ({s}^{-3})$ & $1.8 \times 10^{-35}$ \\\cline{2-3}
        \multicolumn{1}{|l|}{}  & $\ddot{\nu}_{\parallel}\, ({s}^{-3})$ & $\left(-8 \pm 3\right)\times 10^{-30}$ \\\cline{2-3}
        \multicolumn{1}{|l|}{}  & $\ddot{\nu}_{\perp}\, ({s}^{-3})$ & $1.34 \times 10^{-31}$ \\\cline{2-3}
        \multicolumn{1}{|l|}{}  & $\ddot{\nu}_{\rm jerk}\, ({s}^{-3})$ & $7.2 \times 10^{-32}$ \\\hline
        & $\sigma(\ddot{\nu})\, ({s}^{-3})$ & $3.8 \times 10^{-30}$ \\\cline{2-3}
    \end{tabular}
    \caption{Spin parameters and dynamical contributions to $\ddot{\nu}$ for PSR J1909$-$3744, calculated using Eq.~\eqref{eq:nu_ddot:obs}. We used $\ddot{\nu}_i = n \dot{\nu}_i^2 / \nu$ with $n=3$ to estimate the intrinsic value of $\ddot{\nu}_{i}$.}
    \label{tab:psr}
\end{table}

\section{Conclusion}

In this work, we provided an in-depth analysis of BNV effects on pulsar spin-down, examining a broad spectrum of models describable independent of the specific particle physics of BNV. These models lead to a quasi-equilibrium transformation in pulsars, guided by a baryon-conserving EoS. Our study underscores the potential to observe diverse braking indices, both positive and negative, particularly when BNV and spin-down rates are closely matched. We have also delineated the conditions under which ``dead'' pulsars may experience a reactivation due to BNV.

Assuming the BNV rate is predominantly density-driven, it emerges as a universal effect tied solely to a pulsar's mass. If this effect amplifies with increasing density, heavier pulsars, such as PSR J0348+0432—one of the heaviest known—would exhibit higher BNV rates. This observation places an upper limit on the BNV rate, restricting it to $\Gamma \lesssim 10^{-10}\, {\rm yr}^{-1}$. This constraint, however, does not preclude the possibility of non-universal, temperature-dependent BNV scenarios, beckoning further exploration into the complex particle physics underpinning these processes. Moreover, incorporating BNV processes that modify the EoS necessitates a more thorough examination of the particle physics governing BNV. 

Future constraints or detection of BNV effects could capitalize on the precise timing of millisecond pulsars. The current timing precision, sensitive to $\Gamma \sim 10^{-9}\, {\rm yr}^{-1}$, underscores the critical role of enhancing timing methodologies. Increasing the timing precision, expanding the array of observed pulsars, and extending observation durations emerge as crucial for the detection of BNV effects amidst the amalgam of intrinsic and extrinsic pulsar timing noises~\cite{2013CQGra..30v4011L}. 

Importantly, detecting BNV can be complicated by extrinsic effects due to the relative motion of pulsars. Hence, measurements of $\ddot{\nu}$ that consider known extrinsic factors, including established three-dimensional velocities, are considered ideal~\cite{Liu:2020hkx}. Furthermore, an improved understanding of Galactic dynamics and the precise modeling of extrinsic effects will be crucial for accurately attributing observed anomalies to BNV. An intriguing approach lies in harnessing binary pulsar timing to map the structure of the Galaxy. This technique offers the potential to discern the instantaneous accelerations experienced by binary pulsars, thereby offering insights that surpass the limitations of traditional Galactic modeling methodologies~\cite{2024arXiv240115808D}. Conventional models depend on assumed symmetries and equilibrium conditions, potentially overlooking the coarse and non-equilibrium characteristics of the Galactic environment~\cite{10.1093/mnras/136.1.101}. 

As we stand on the cusp of a new era in pulsar astronomy, the synergy between enhanced observational techniques and deeper theoretical insights into pulsars may potentially reshape our understanding of the fundamental symmetries that govern our universe by uncovering the elusive signatures of baryon number violation.

\acknowledgments
I acknowledge support from the U.S. Department of Energy under contract DE-FG02-96ER40989. My thanks go to Jeffrey Berryman, Susan Gardner, and David Tsang for their insightful discussions, which have contributed to this work. I am also grateful to the Center for Theoretical Underground Physics and Related Areas (CETUP*), The Institute for Underground Science at Sanford Underground Research Facility (SURF), and the South Dakota Science and Technology Authority for their hospitality and financial support. Their stimulating environment was invaluable during the period in which part of this work was completed. I extend special thanks to Dale and Tina Carter for their exceptional hospitality during my stay in Lead, South Dakota.

\bibliographystyle{JHEP}
\bibliography{PTA_BNV}

\end{document}